\documentclass[pra,twocolumn,showpacs,preprintnumbers,amsmath,amssymb, superscriptaddress]{revtex4-1}

\usepackage{graphicx}% Include figure files
\usepackage{dcolumn}% Align table columns on decimal point
\usepackage{bm}% bold math
\usepackage{epsfig}
\usepackage{amsmath}
\usepackage{amssymb}
\usepackage{color}
\usepackage{bbm}
\usepackage[caption=false]{subfig}
\usepackage{placeins}
\usepackage{soul}

\usepackage[colorlinks=true,citecolor=blue,linkcolor=blue,urlcolor=blue]{hyperref}
\usepackage{cancel}

\newcommand{\ket}[1]{|#1\rangle}

\newcommand{\erw}[1]{\langle#1\rangle}
\newcommand{\erwa}[1]{\left\langle#1\right\rangle}
\newcommand{\abs}[1]{\lvert#1\rvert}

\def\equationautorefname~#1\null{%
  Eq.~#1\null
}
\def\figureautorefname~#1\null{%
	Fig.~#1\null
}
\graphicspath{{./figs/}}

\begin{document}

\title{Non-equilibrium magnetic phases in spin lattices with gain and loss}

\author{Julian Huber}
%\email{julian.huber@ati.ac.at}
\affiliation{Vienna Center for Quantum Science and Technology,
Atominstitut, TU Wien, 1040 Vienna, Austria}
\author{Peter Kirton}
\affiliation{Vienna Center for Quantum Science and Technology,
	Atominstitut, TU Wien, 1040 Vienna, Austria}
\affiliation{Department of Physics and SUPA, University of Strathclyde, Glasgow G4 0NG, UK}
\author{Peter Rabl}
\affiliation{Vienna Center for Quantum Science and Technology,
Atominstitut, TU Wien, 1040 Vienna, Austria}

\date{\today}

%%%%
\begin{abstract}
We study the magnetic phases of a non-equilibrium spin chain, where coherent interactions between neighboring lattice sites compete with alternating gain and loss processes. This competition between coherent and incoherent dynamics induces transitions between magnetically aligned and highly mixed phases, across which the system changes from a low- to an effective infinite-temperature state. We show that the origin of these transitions can be traced back to the dynamical effect of parity-time-reversal symmetry breaking, which has no counterpart in the theory of equilibrium phase transitions. This mechanism also results in very atypical features and we find first-order transitions without phase co-existence and mixed-order transitions which do not break the underlying $U(1)$ symmetry, even in the appropriate thermodynamic limit. Thus, despite its simplicity, the current model considerably extends the phenomenology of non-equilibrium phase transitions beyond that commonly assumed for driven-dissipative spins and related systems.
 \end{abstract}

\maketitle
\section{Introduction}
Magnetically ordered and disordered phases are ubiquitous in interacting spin systems and represent an area of intensive research in condensed-matter and statistical physics. Such phenomena are studied in thermal equilibrium where, for example, order-favoring interactions compete  with thermal or quantum fluctuations. A topic of growing interest is the study of non-equilibrium properties of interacting spins or other quantum many-body systems, in particular, in the presence of external driving and dissipation~\cite{Walls1978,Dimer2007,Morrison2008,Prosen2008,Diehl2008,Kessler2012,Lee2013,Sieberer2013,Zou2014,Marzolino2014,Carmichael2015, Weimer2015,Schiro2016,Rossini2016,Maghrebi2016,Buchold2017, Domokos2017,FossFeig2017,Orus2017,Rota2017,Ciuti2018,Jin2018,Minganti2018,Hannukainen2018,Roscher2018,Kirton2019,Gillman2019,Vicentini2019,Ferreira2019,Barberena2019,Puel2019,Verstraelen2020,Landa2020}. Such conditions are naturally found  in quantum optical and cold atom settings~\cite{Syassen2008,Baumann2010,Barreiro2011,Muller2012,Safavi-Naini2018,Lienhard2018,Wade2018}. In these systems, for example, trapped atoms are highly isolated from the environment, while efficient dissipation channels can be engineered through optical pumping and laser cooling techniques. However, in contrast to their equilibrium counterparts, the stationary states of such systems are no longer determined by energetic considerations or by the minimization of a thermodynamic potential. As a consequence, there is still little known about the general principles that govern the formation and the properties of ordered and disordered phases of such driven-dissipative quantum systems.

In the context of spin systems, a lot of previous work on this topic has been focused on the effect of dissipation on the stationary phases of the transverse field Ising and related XYZ models~\cite{Prosen2008,Lee2013,Weimer2015,Rossini2016,Rota2017,Jin2018,Puel2019,Landa2020}. While the equilibrium properties of such models are well-known, a general problem in the study of their dissipative counterparts is that reliable numerical simulations are only available in one dimension (1D), where due to the added damping and (non-equilibrium) fluctuations, typically no sharp transitions occur~\cite{Rossini2016,Landa2020}. Notably exceptions to this rule are certain classes of boundary-driven spin models, where dissipation only occurs at the ends~\cite{Prosen2008,Marzolino2014,Puel2019}.    In 2D and higher dimensions, where phase transitions are more easily engineered, exact numerical simulations are restricted to rather small lattices~\cite{Rossini2016,Rota2017,Jin2018,Landa2020}, while predictions from mean-field theory are still questionable. Therefore, most of our more reliable insights about dissipative phase transitions are currently based on studies of zero-dimensional models, involving, for example, a collective spin $S$ system~\cite{Morrison2008,Kessler2012,Ferreira2019,Barberena2019,Walls1978,Hannukainen2018}, a weakly nonlinear bosonic mode~\cite{Drummond1980,Casteels2017,Bartolo2016} or combinations of both~\cite{Dimer2007,Domokos2017,Kirton2019}.  In this case sharp phase transitions can appear for $S\rightarrow \infty$ or equivalent semiclassical limits. The steady states of such models can be calculated numerically for sufficiently large system sizes and although these systems exhibit phases with enhanced fluctuations,  mean-field theory and linearization techniques typically still provide a very accurate qualitative description. From the analysis of many such systems a common picture of dissipative phase transitions emerged~\cite{Kessler2012,Minganti2018}, where---in essence---energy gaps are replaced by dissipation rates, but where the actual  phenomenology is still very similar to the equilibrium case:  There are discontinuous first-order phase transitions near which two distinct quasi-stationary states can coexist and continuous second-order phase transitions associated with the breaking of a  symmetry. Naturally, this motivates the search for non-equilibrium critical phenomena that lie outside of this conventional framework and for the basic mechanisms that may cause such behavior.

\begin{figure}
	\includegraphics[width=\columnwidth]{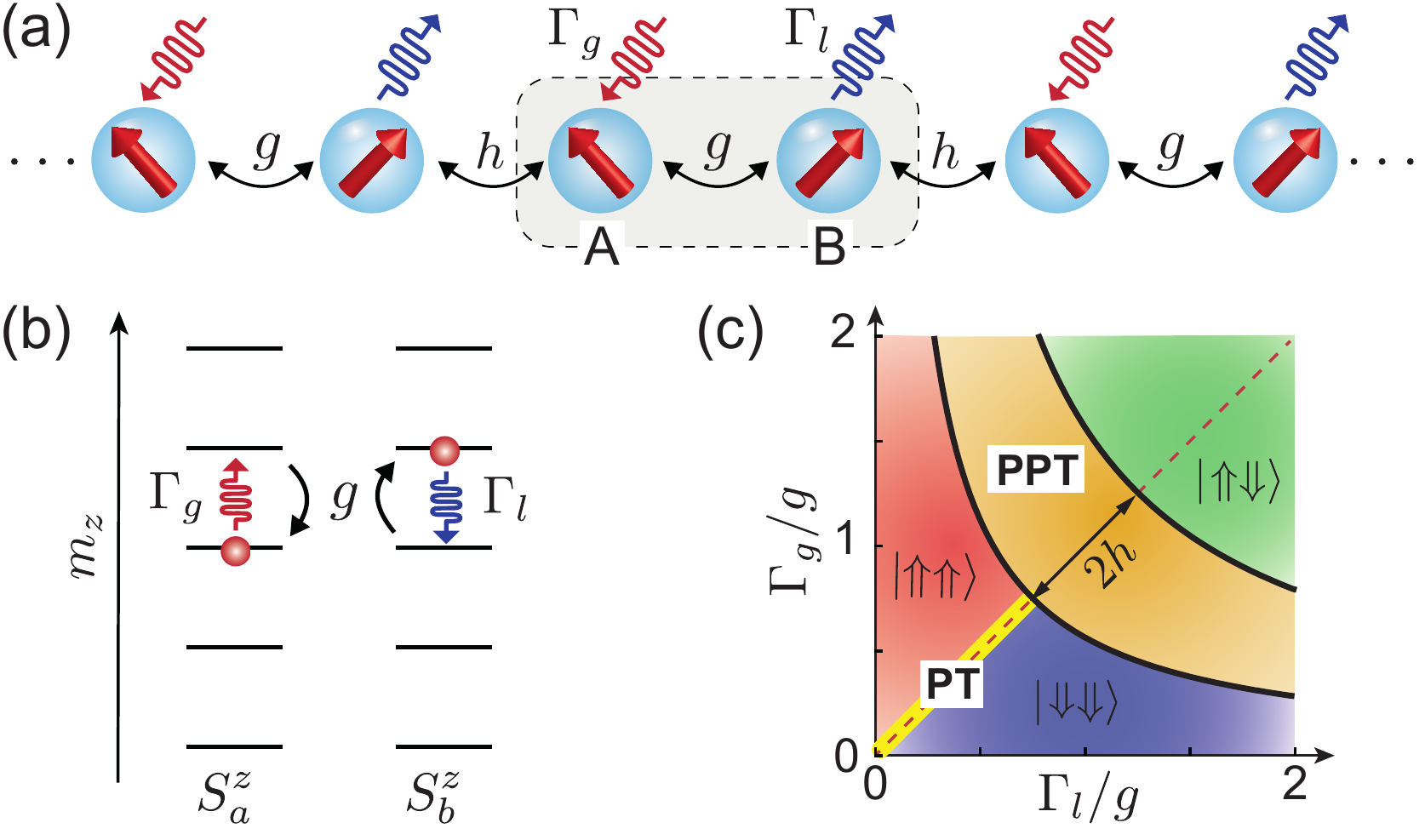}
	\caption{(a) Sketch of a 1D spin chain, where the individual spins are coherently coupled to their neighbors and  alternately pumped with rate $\Gamma_{g}$ or cooled with rate $\Gamma_{l}$. (b) Illustration of the coherent and dissipative processes within a single unit cell. (c) Plot of the steady-state phase diagram of the dissipative spin chain as a function of the gain and loss rates. The solid lines indicate the phase boundaries for $S\to\infty$. }
	\label{Fig1:Setup}
\end{figure} 

In this paper we propose and analyze a minimal lattice spin model as depicted in Fig.~\ref{Fig1:Setup}(a) for studying non-equilibrium phenomena that go beyond the picture discussed above. In this setting, neighboring spins in a large bias field are coupled via excitation-conserving XX interactions, such that the ground state of the system is always a trivial paramagnet. This allows us, first of all, to investigate emergent magnetic phases that do not exist in equilibrium and are solely induced by the addition of incoherent processes in form of alternating gain and loss. In the following analysis we show that this simple model already exhibits several transitions between magnetically-aligned and strongly mixed states, which do not exhibit the usual phenomenology of first- and second-order phase transitions. Specifically, we find first-order transitions without phase co-existence and mixed-order transitions, where even in the limit of large spin quantum numbers the underlying $U(1)$ symmetry of the model is not broken. This is in stark contrast to what is obtained from mean-field predictions~\cite{Lee2013}, which are expected to be very accurate in this limit, but also from more general considerations about phase transitions in Liouvillian systems~\cite{Minganti2018}. We show that this qualitative discrepancy  can be explained by the  mechanism of PT (parity and time-reversal) symmetry breaking~\cite{Bender1998,Ganainy2018}, which is mainly known from the dynamics of  (classical) non-Hermitian systems with balanced gain and loss. Our analysis demonstrates that this transient dynamical effect, which has no counterpart in equilibrium or  isolated quantum systems, also determines the stationary state properties. Interestingly, in extended lattice systems this is still the case even when the respective Liouvillian symmetry~\cite{Huber2020} is not exactly fulfilled. Therefore, beyond the specific model considered here, this insight will also be important to characterize and classify non-equilibrium phases in many other models or higher-dimensional lattice geometries, where mean-field theory can fail and exact numerical simulations are not available.

\section{Model}
We consider a one dimensional (1D) chain of $2N$ spin-$S$ systems, which is divided into two sublattices A and B [cf.~Fig.~\ref{Fig1:Setup}(a)]. The spins precess around a static field along the $z$-direction with Larmor frequency $\omega_0$ and are coupled to their neighbors via spin-flip interactions with  alternating strengths $g$ and $h$. The coherent dynamics of this system is described by the Heisenberg model $\mathcal{H}=\hbar \omega_0 M_z+\mathcal{H}_{XX}$, where $M_z=\sum_{n=1}^N (S_{a,n}^z+S_{b,n}^z)$ is the total magnetization and
\begin{equation}\label{eq:H}
\mathcal{H}_{XX}= \frac{\hbar}{2S}\sum_{n=1}^{N} \left(g S_{a,n}^+ S_{b,n}^- +h S_{b,n}^+ S_{a,n+1}^- + {\rm H. c.} \right).
\end{equation}
The $S^{k}_{a,n}$ and $S^{k}_{b,n}$, with $k\in\{x,y,z,\pm\}$, denote the usual spin operators for sublattices $A$ and $B$. Within the parameter regime of interest, $\omega_0\gg g,h$, this model only has a trivial, fully polarized ground state, which would be stabilized by adding decay for all spins. To obtain non-trivial dissipation effects, we thus consider alternately pumping the spins along opposite directions. By changing into a frame rotating with $\omega_0$, the resulting evolution of the system density operator $\rho$ is then described by the master equation (ME),
\begin{equation}\label{eq:ME}
\dot{\rho}=\frac{i}{\hbar} [\rho, \mathcal{H}_{XX}]+  \frac{1}{2S}\sum_{n=1}^N \left(\Gamma_{g} \mathcal{D}[S_{a,n}^+] + \Gamma_{l} \mathcal{D}[S_{b,n}^-]\right)\rho,
\end{equation}
where  $\mathcal{D}[S^{\pm}]\rho=\left(2 S^\pm\rho S^{\mp} - S^{\mp} S^{\pm} \rho - \rho S^{\mp} S^{\pm}\right)$ and $\Gamma_g$ and $\Gamma_l$ are the gain and loss rates, respectively. 
In Eqs.~\eqref{eq:H} and~\eqref{eq:ME} the couplings and pumping rates are scaled by the spin quantum number $S$ to ensure that the relevant timescales of the system dynamics remain the same for different total spin. Note that Eq.~\eqref{eq:ME} preserves the $U(1)$ symmetry associated with a common rotation of all the spins in the $x$--$y$ plane. In Sec.~\ref{sec:Implementation} below we discuss possible experimental implementations of this model using, for example, ensembles of cold atoms or solid-state defects in coupled cavity arrays. 

As depicted in Fig.~\ref{Fig1:Setup}(b), the dissipative terms in Eq.~\eqref{eq:ME} drive the system into a state with a staggered magnetization, while the coherent coupling tends to counteract this imbalance. This competition leads to several distinct phases for the steady state of the spin chain, $\rho_{0}=\rho(t\rightarrow\infty)$,  which are summarized in Fig.~\ref{Fig1:Setup}(c). We identify two types of ordered phases, which exhibit either anti-ferromagnetic (AM) or ferromagnetic (FM) alignment of the spins. In addition, there are two strongly disordered phases, which are labeled as PT-symmetric (PT) and pseudo-PT-symmetric (PPT) for reasons that will be discussed in more detail below. In the limit $S\to\infty$ the five different phases are separated by sharp boundaries defined by the lines 
\begin{equation}\label{eq:PhaseBoundary}
\Gamma_{g} \Gamma_{l}=(g\pm h)^2
\end{equation}
%$\Gamma_{g} \Gamma_{l}=(g\pm h)^2$  
and $\Gamma_{g}=\Gamma_{l}$ for $\Gamma_{g} \Gamma_{l}<(g- h)^2$, which can be derived from a Holstein-Primakoff approximation (HPA) (see Appendix~\ref{app:HPApproximation}).

\section{Dissipative spin dimer}\label{sec:Dimer}
\begin{figure}
	\includegraphics[width=\columnwidth]{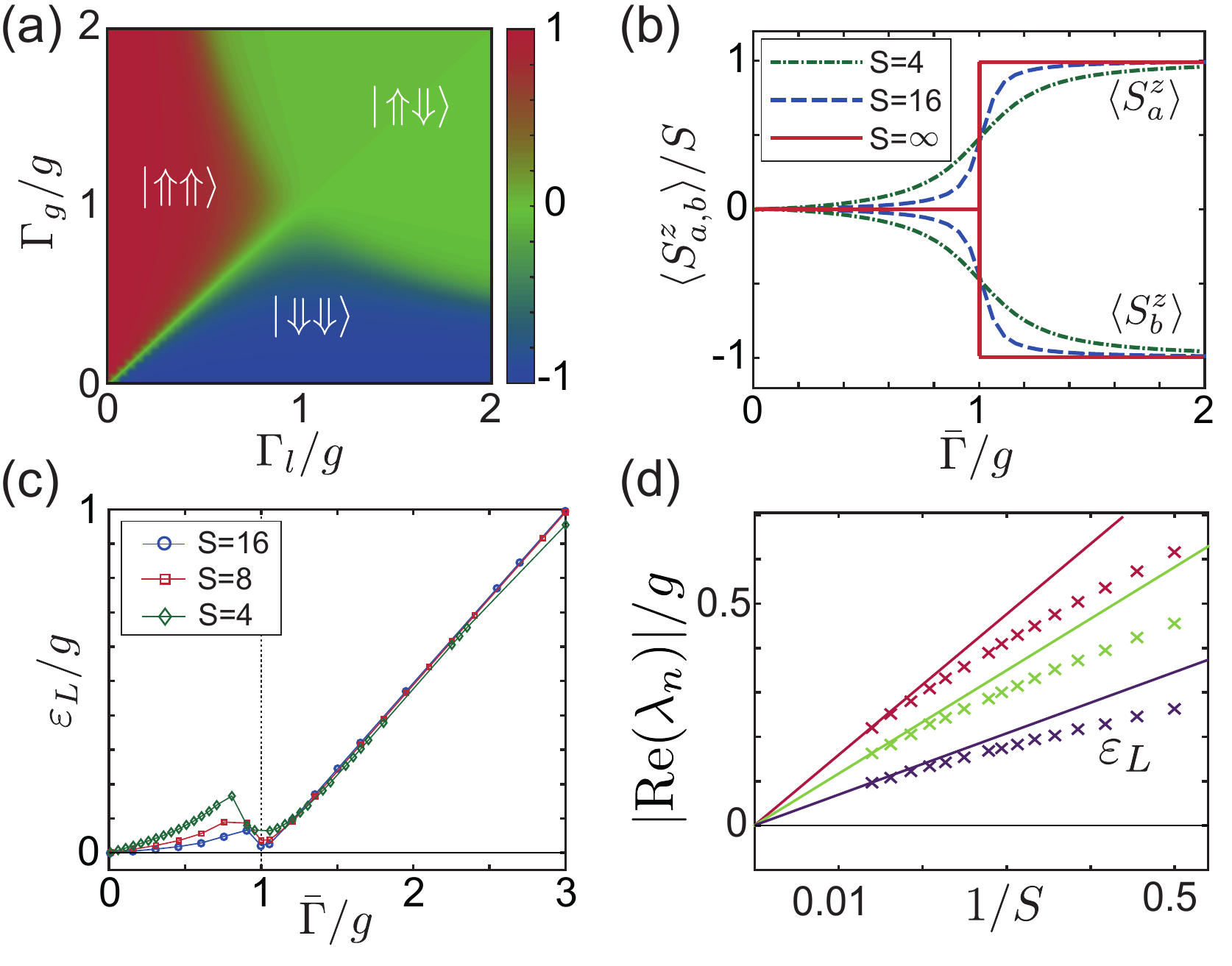}
		\caption{(a) Plot of the order parameter $\mathcal{M}_z=\langle M_z\rangle/(2S)$ for a spin dimer with $S=12$.  (b) Magnetization of the individual spins along the symmetry line, $\Gamma_l=\Gamma_g=\bar \Gamma$, for different spin quantum numbers. For the same parameters, (c) shows the dependence of the Liouvillian gap, $\varepsilon_L$, on the ratio $\bar \Gamma/g$. (d) Scaling of $\varepsilon_L$ and of two additional eigenvalues at the transition point, $\bar\Gamma=g$. The crosses are the exact numerical results for up to {\bf $S=18$} and the solid lines are linear extrapolations to zero, i.e., $\varepsilon_L\sim 1/S$.}
	\label{Fig2:Dimer}
\end{figure}

To understand some basic properties of the model, it is instructive to first consider the limit $h\rightarrow 0$, where the chain separates into decoupled spin dimers. In this case the intermediate mixed phase disappears and for the remaining phases the order parameter $\mathcal{M}_z=\langle M_z\rangle/(2S)$ is shown in Fig.~\ref{Fig2:Dimer}(a). For $\Gamma_{g,l}\gg g$ dissipation always dominates and the spins are simply pumped into an anti-aligned AM configuration, where $\mathcal{M}_z\approx 0$, but $\langle S^z_a\rangle=-\langle S_b^z\rangle\approx S$. For $\sqrt{\Gamma_g \Gamma_{l}}<g$, this arrangement is destabilized by the coherent coupling, which, in this regime, efficiently redistributes energy between the two sites. As a result, the stationary state is only determined by the sign of the net damping rate, $\delta \Gamma=(\Gamma_g-\Gamma_l)$, and exhibits FM alignment, $\mathcal{M}_z \simeq \pm 1$. This ordered phase extends into the regime $\Gamma_{g,l}\ll g$, where the coherent interaction dominates and where one would thus expect a highly mixed, depolarized phase. At the same time the spin alignment opposes the applied dissipation in one of the sublattices, which shows that this type of order still depends on a non-trivial interplay between coherent and incoherent processes. Interestingly, even for $S\gg 1$ this stationary ferromagnetic alignment is not captured by the mean-field equations of motion (see Appendix~\ref{app:MF}), which instead predict a limit cycle for one of the spins with a vanishing average magnetization. 

\subsection{PT symmetry}
Of specific interest is the behavior of this system along the diagonal $ \Gamma_l=\Gamma_g$, which for $\bar \Gamma=(\Gamma_g+\Gamma_l)/2 < g$ marks the boundary between the two FM phases. Along this line the model becomes PT symmetric~\cite{Huber2020}. This means that the ME, Eq.~\eqref{eq:ME}, is invariant under the combined  exchange of sublattices A and B (parity)  and the conjugation of the jump operators $S^+\leftrightarrow S^{-}$ (exchanging loss and gain, i.e., reversing time). Very generically, the existence of this symmetry, which is only defined for dissipative systems, implies that the steady state for $\bar \Gamma\ll g$ is close to the (symmetric) fully mixed state~\cite{Huber2020},
\begin{equation}\label{eq:rho0}
\rho_0 \simeq  \frac{1}{(2S+1)^2} \left[ \mathbbm{1}+ O \left(\frac{\bar \Gamma}{g}\right)\right],
%\frac{\bar \Gamma^2}{g (2S+1)} \left(S_a^z-S_b^z\right)\right],
\end{equation}
with $\langle M_z\rangle\simeq \langle S^z_{a,b}\rangle\simeq 0$, and that this phase is separated from the (symmetry-broken)  AM phase by a sharp transition in the limit $S\rightarrow \infty$. This behavior is clearly visible in Fig.~\ref{Fig2:Dimer}(b), where we plot the individual magnetizations along the line $\Gamma_l=\Gamma_g$ for increasing $S$.

In Fig.~\ref{Fig2:Dimer}(c) and (d) we also plot the real part of the smallest magnitude non-zero eigenvalues, $\lambda_n$, of the Liouville superoperator $\mathcal{L}$, which is defined by $\dot \rho=\mathcal{L}\rho$.  As we approach the transition point $\bar \Gamma=g$ from the AM phase, we observe a closing of the Liouvillian gap, $\varepsilon_L\sim 1/S$, where $\varepsilon_L={\rm min}\{-{\rm Re}(\lambda_n)\}$. While the closing of the Liouvillian gap is expected for any dissipative phase transition point~\cite{Kessler2012,Minganti2018}, we also find that many of the larger magnitude eigenvalues of $\mathcal{L}$ vanish and remain vanishingly small (in the limit of large $S$) within the whole PT phase. This indicates that for $\bar \Gamma<g$ the gain and loss processes cancel out on average. In contrast, fluctuations, which still occur with rates $\Gamma_{g,l}$, are not reduced correspondingly and drive the system into a highly mixed state. Since the energy levels of the system do not change at the transition point, this sudden increase of entropy translates into a jump of the systems' effective temperature~\cite{Kessler2012, Roscher2018,Hannukainen2018}. This is a crucial difference to equilibrium systems, where the level of fluctuations is determined by a fixed temperature in all  phases. 

More specifically, as already pointed out in Eq.~\eqref{eq:rho0} above, the steady state in this PT symmetric phase is close to the fully mixed, i.e., infinite temperature state. This must be contrasted to states with a high, but finite temperature as observed in other models~\cite{Kessler2012,Roscher2018}, since the impurity of the system, $\mathcal{I}=1/\mathcal{P}$, becomes extensive,  
\begin{equation}
\lim_{S\rightarrow \infty} \frac{\mathcal{I}(\delta \Gamma=0)}{(2S+1)^2}>0.
\end{equation}
This implies that such a state cannot be approximated by a mean-field ansatz, since for any observable fluctuations dominate over its mean value. As we will discuss in the following, many of the unusual features of the current model can be traced back to this specific property of the PT symmetric phase. Note that a similar transition between low and infinite temperature phases can also occur in various other models~\cite{Hannukainen2018,Barberena2019}.
% {\color{red} cite models, with transition into infinite temperature phase, even of the purity is not explicitly discussed}.
It is thus important to develop a more general understanding of this type of transition, in particular in extended lattice systems, where the fate of such infinite temperatures phases is still unknown.

\subsection{Absence of phase co-existence}
For the dimer model, Fig.~\ref{Fig2:Dimer}(a) shows that all transitions are of first order, meaning that at the respective transition lines the magnetization in the limit $S\rightarrow \infty$ jumps abruptly between two different values. For concreteness, we will focus in the following on the transition between the two FM phases for $\bar \Gamma <g$. This situation is reminiscent of a regular Ising ferromagnet in the presence of an external bias field $B$, a role which is here taken by the rate imbalance $\delta \Gamma$. In an equilibrium magnetic system and for $B=0$, there is no externally imposed direction and the magnetic moments then spontaneously align themselves along one of the two possible directions. When averaged over these two equally probable configurations, the resulting density operator corresponds to an equal mixture between the two FM states. 

\begin{figure}
	\includegraphics[width=\columnwidth]{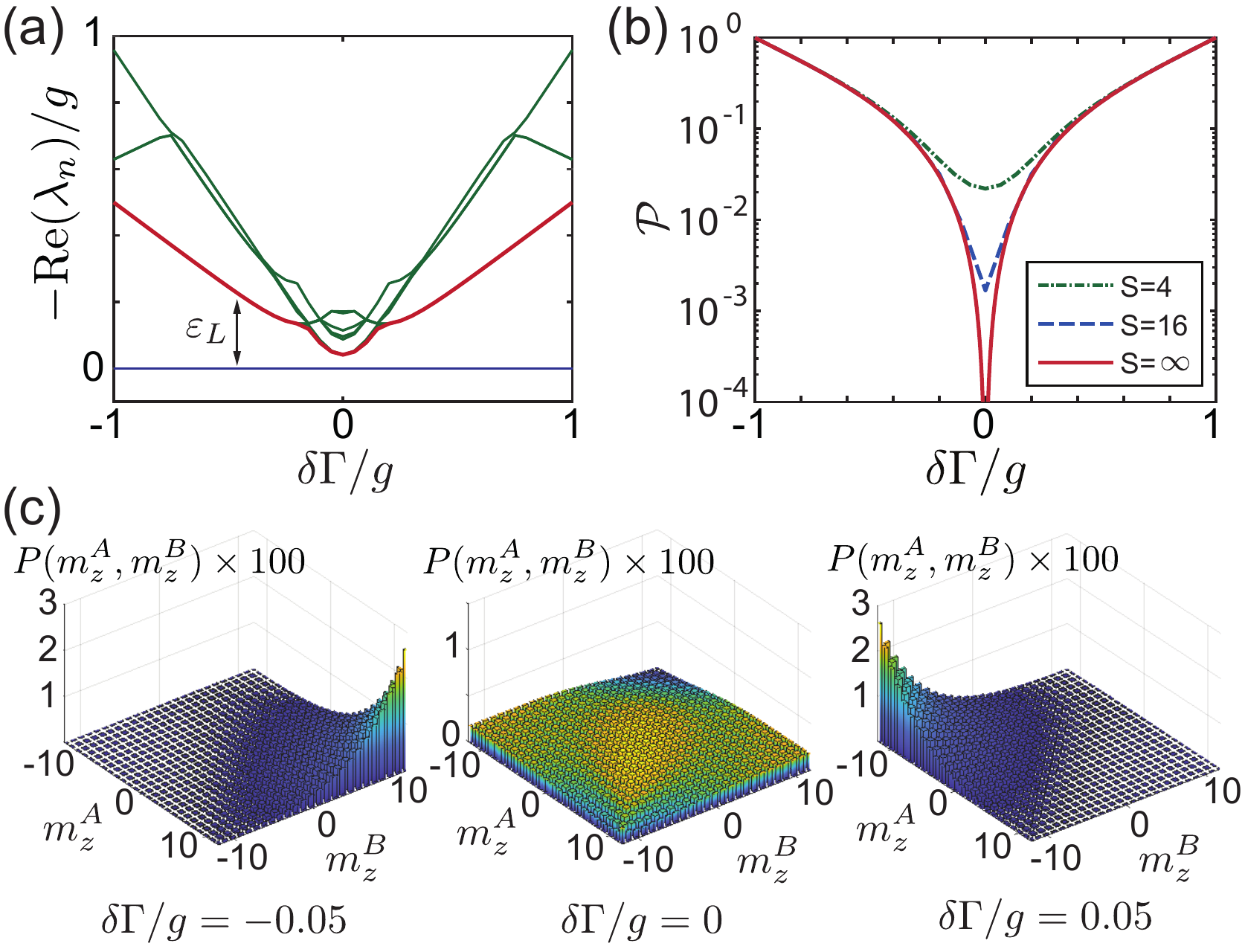}
	\caption{(a) The real part of the first 8 eigenvalues $\lambda_n$ of the Liouvillian $\mathcal{L}$ for a fixed value $\bar{\Gamma}=0.5g$ and $S=16$. In the limit $S\rightarrow\infty$, the point $\delta \Gamma=0$ marks the phase transition line between the two FM states. (b) Purity of the steady-state, $\mathcal{P}={\rm Tr}\{ \rho_0^2\}$, for the same parameters but different values of $S$. The line labelled $S=\infty$ shows the analytic prediction obtained from a HPA (see Appendix~\ref{app:HPApproximation}). (c) The probability distribution $P(m_z^A,m_z^B)$ for the magnetization values of each spin is shown for three different values of $\delta\Gamma$ representing the steady state just below, at, and just above the transition point for $S=12$.}
	\label{Fig3:DimerGL}
\end{figure}

It has been previously conjectured~\cite{Minganti2018} that such a picture should also apply, very generically, to discontinuous  transitions in driven-dissipative systems. This conclusion is primarily based on the analysis of the dissipative Kerr-oscillator (see discussion below), where this analogy between equilibrium and non-equilibrium phase transitions is indeed very accurate. However, the current model demonstrates that there are other types of first-order phase transitions, where this analogy does not apply. To illustrate this point we study in Fig.~\ref{Fig3:DimerGL} in more detail the behavior of the system as we tune it across the transition line for a fixed $\bar \Gamma/g=0.5$ and varying $\delta \Gamma$. First of all, Fig.~\ref{Fig3:DimerGL}(a) shows the expected closing of the Liouvillian gap at $\delta \Gamma=0$ confirming the existence of a sharp phase transition in the limit $S\rightarrow 0$ [see also Fig.~\ref{Fig2:Dimer}(c)]. In Fig.~\ref{Fig3:DimerGL}(b) we plot the purity of the steady state, which vanishes as $\mathcal{P}\sim 1/(2S+1)^2$ at the transition point. More explicitly, Fig.~\ref{Fig3:DimerGL}(c) shows the probability distribution $P(m_z^A,m_z^B)$ for the magnetization values of each spin just below, at and just above the transition point. This comparison demonstrates that the state at $\delta \Gamma=0$ is clearly different from a naively expected mixture between the two neighboring phases. Although in the middle plot we still see some small variations in $P(m_z^A,m_z^B)$, the scaled impurity in this (finite $S$) example reaches a value of $\mathcal{I}/(2S+1)^2\simeq 0.957$. This confirms that for $S\gg1$ the system transitions between the two opposite FM configurations via an intermediate, fully mixed phase. 

It is instructive to contrast the behavior in Fig.~\ref{Fig3:DimerGL} with the regular first-order phase transition in the dissipative Kerr oscillator mentioned above. The Kerr oscillator is a single nonlinear bosonic mode with annihilation operator $c$, which is described by the Hamiltonian~\cite{Drummond1980,Bartolo2016,Casteels2017}
\begin{equation}
\mathcal{H}_{\rm K}=-\hbar \Delta c^\dagger c + \hbar \frac{U}{D} c^\dagger c^\dagger c c + \hbar \sqrt{D}  F (c^\dagger+c).
\end{equation}
Here $U$ is the strength of the nonlinearity and $F$ the strength of an external driving field, which is detuned from resonance by $\Delta$. The parameter $D$ plays the role of an effective Hilbert space dimension such that $D\rightarrow \infty$ represents the thermodynamic limit of this model. The dynamics of the dissipative Kerr oscillator is then described by the ME   
\begin{equation}\label{eq:LKerr}
\dot{\rho} = -\frac{i}{\hbar} [\mathcal{H}_{\rm K},\rho]+\gamma \mathcal{D}[c] \equiv \mathcal{L}_{\rm K}\rho,
\end{equation}
where  $\gamma$ is the decay rate. The steady state of this ME exhibits a first-order phase transition at $F/\gamma\simeq 1.76$, where the system switches between states with a low and high photon number expectation value. 

\begin{figure}
	\includegraphics[width=\columnwidth]{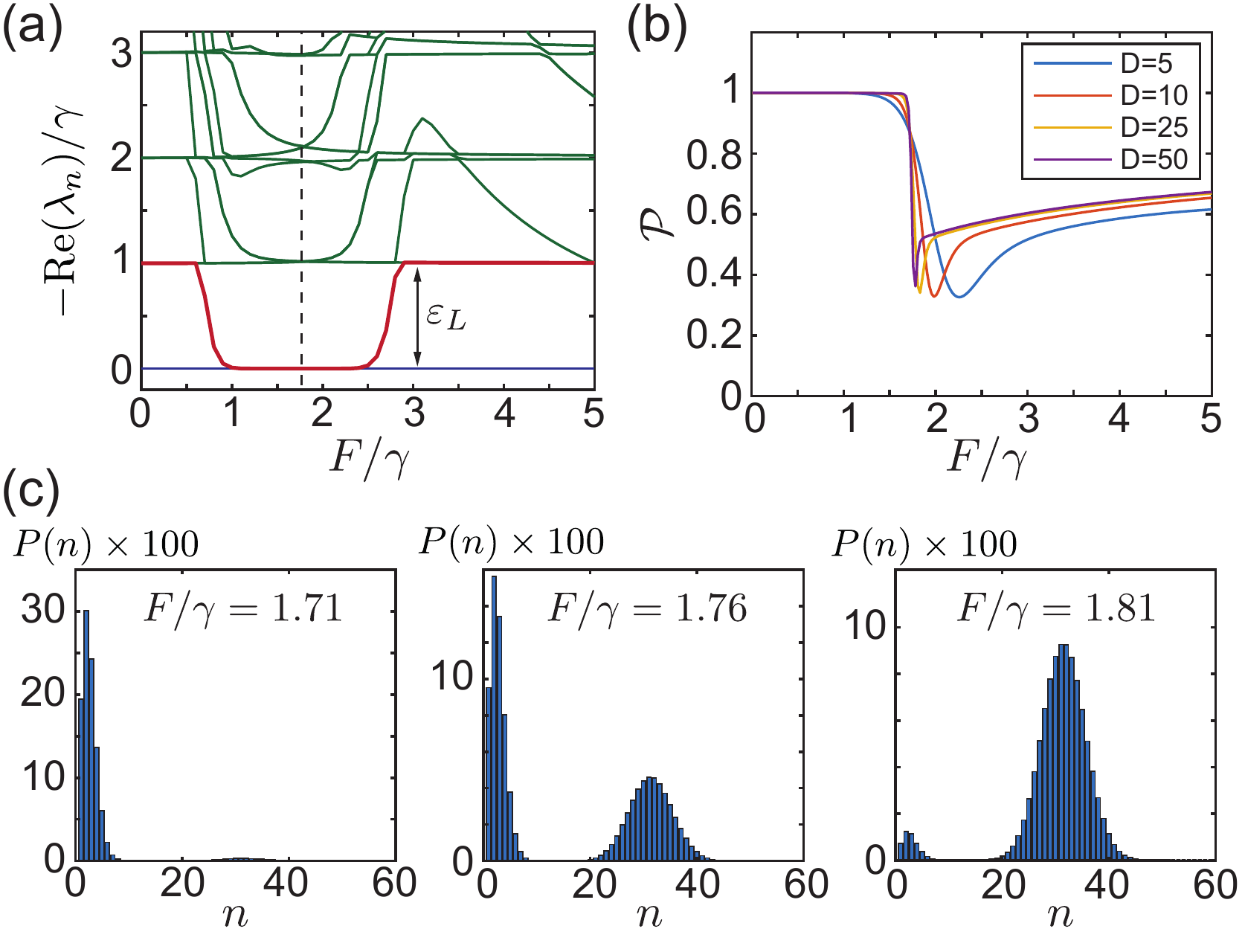}
	\caption{First-order phase transition in the disspative Kerr oscillator as defined in Eq.~\eqref{eq:LKerr}. (a) The real part of smallest eigenvalues, $\lambda_n$, of the Liouvillian $\mathcal{L}_{\rm K}$ as a function of $F/\gamma$ and for $U/\gamma=10$, $\Delta/\gamma=10$ and $D=50$. The dashed vertical line marks the phase transition point at $F/\gamma\simeq 1.76$. (b) Purity  of the steady-state $\rho_0$, where $\mathcal{L}_{\rm K}\rho_0=0$, for the same parameters but different values of $D$. (c) Probability distribution $P(n)$ for the oscillator number states $|n\rangle$ just below, at, and just above the transition point  and for $D=50$.}
	\label{Fig4:KerrOsc}
\end{figure}

Figure~\ref{Fig4:KerrOsc} summarizes the behavior of the Kerr oscillator when it is tuned across this transition point, which we can contrast with the observations in Fig.~\ref{Fig3:DimerGL}. We first notice that the Liouvillian gap is vanishingly small over a larger parameter range and it vanishes as $\varepsilon_L\sim e^{-D}$ at the transition point \cite{Casteels2017}. In contrast to the spin model, only two eigenvalues vanish, which already indicates that at the transition point the system is well described by a mixture of two distinct metastable states. This picture is also confirmed by a non-vanishing purity in Fig.~\ref{Fig4:KerrOsc}(b) and the distribution of the occupation numbers of the oscillator states, $p(n)$, in Fig.~\ref{Fig4:KerrOsc}(c). This last result clearly shows that the state at the transition point is a mixture of the two neighboring phases, which can also be verified explicitly~\cite{Minganti2018}. 

The observation that such a co-existence between the two FM states does not occur for the spin dimer can be attributed to the fact that in this model a large number of Liouvillian eigenvalues vanish at the same time near $\delta\Gamma=0$. This provides, roughly speaking, more flexibility to construct the actual steady state out of many nearly-degenerate eigenvectors of $\mathcal{L}$. Since in the spin model the closing of the Liouvillian gap only scales inversely with the system size and not exponentially also means that other properties, such as the divergence of the relaxation rate, etc., will be very different in these two types of first-order transition.

%To investigate the impact of this unconventional behavior on the phases and correlations of extended spin systems, 

\section{Dissipative spin chain}
We now return to the fully coupled chain with $0<h\leq g$ to see how the basic effects discussed above affect the non-equilibrium states of the extended spin lattice. As already mentioned in the introduction, for small spins, $S\sim O(1)$, there are typically no sharp phase transitions in dissipative spin systems in 1D, even for an infinite number of lattice sites $N\rightarrow \infty$. This can be understood from the fact that the fluctuations introduced by the dissipation processes act as a finite effective temperature, which typically prevents long-range order in 1D~\cite{MerminWagner}. Therefore, in the following analysis we retain our focus on the regime $S\gg 1$, as above. While in this limit sharp transitions already occur for a single cell, the resulting phases and the nature of the phase transitions can be very different in the lattice system. In fact, the exact nature of a phase transition can only be determined in extended systems, where, apart from the order parameter, also information about spatial correlations and their critical scaling is available.

\subsection{Simulation of dissipative spin lattices}
While in 1D the dynamics and steady states of dissipative systems with a small local Hilbert space dimension can still be simulated efficiently using matrix product operator techniques~\cite{Vidal2007,Orus2008}, this is not possible for the current system when $S\gg 1$. At the same time, we have seen that, even in the limit of a large spin quantum number, fluctuations are dominant, which makes a mean-field approximation unreliable. To overcome these limitations we developed a stochastic method based on a variant of the truncated Wigner approximation (TWA)~\cite{Olsen2005,Polkovnikov2010,Ng2011,Ng2013} to simulate the phase space distribution of the spins. 
The basic idea of this approach is  to map each of the spins onto two independent bosonic modes $a$ and $b$, by using the Schwinger boson representation
\begin{equation}
	S^{+}=a^\dagger b,\qquad S^{-}=a b^\dagger,\qquad
	S^{z}=\frac{1}{2} (a^\dagger a-b^\dagger b).
\end{equation}
The resulting master equation for the lattice of $4N$ bosonic modes can then be converted into an equivalent partial differential equation for the Wigner function of those modes. The usual TWA  corresponds to neglecting all third and higher order derivatives to obtain a Fokker-Planck equation (FPE). However, this is not enough since in general the diffusion matrix of this FPE is not positive and the distribution cannot be simulated efficiently via stochastic methods. While this problem could be overcome by using the positive-P representation~\cite{Olsen2005,GardinerZoller} instead, this approach still suffers from the appearance of ``spikes", where individual trajectories diverge~\cite{GardinerZoller,Ng2011,Ng2013} and limit stochastic simulations to very short times.

In order to make the TWA applicable for the simulation of the long-time behavior of large spin lattices, we perform an additional \emph{positive diffusion approximation}, where the non-positive terms in the diffusion matrix are also neglected. 
Although only applicable for very large spins, this method goes beyond mean-field theory by accounting  for the relevant quantum noise terms and allows us to simulate the steady states of dissipative spin systems with  $N\sim100$ unit cells.  In the ordered phases, these numerical results are in perfect agreement with analytic predictions based on the HPA~\cite{HolsteinPrimakoff}, as detailed in Appendix~\ref{app:HPApproximation}. In addition, we use infinite matrix product operator (iMPO)~\cite{Vidal2007,Orus2008} and cluster-mean field (CMF) simulations to verify that the main characteristics of the different phases are still present in the limit of small and moderate spin quantum numbers. A detailed derivation of the TWA scheme and its applicability for the simulation of collective spin models is presented in a separate publication~\cite{Huber2021}.

\subsection{The PPT phase}

\begin{figure}
	\includegraphics[width=\columnwidth]{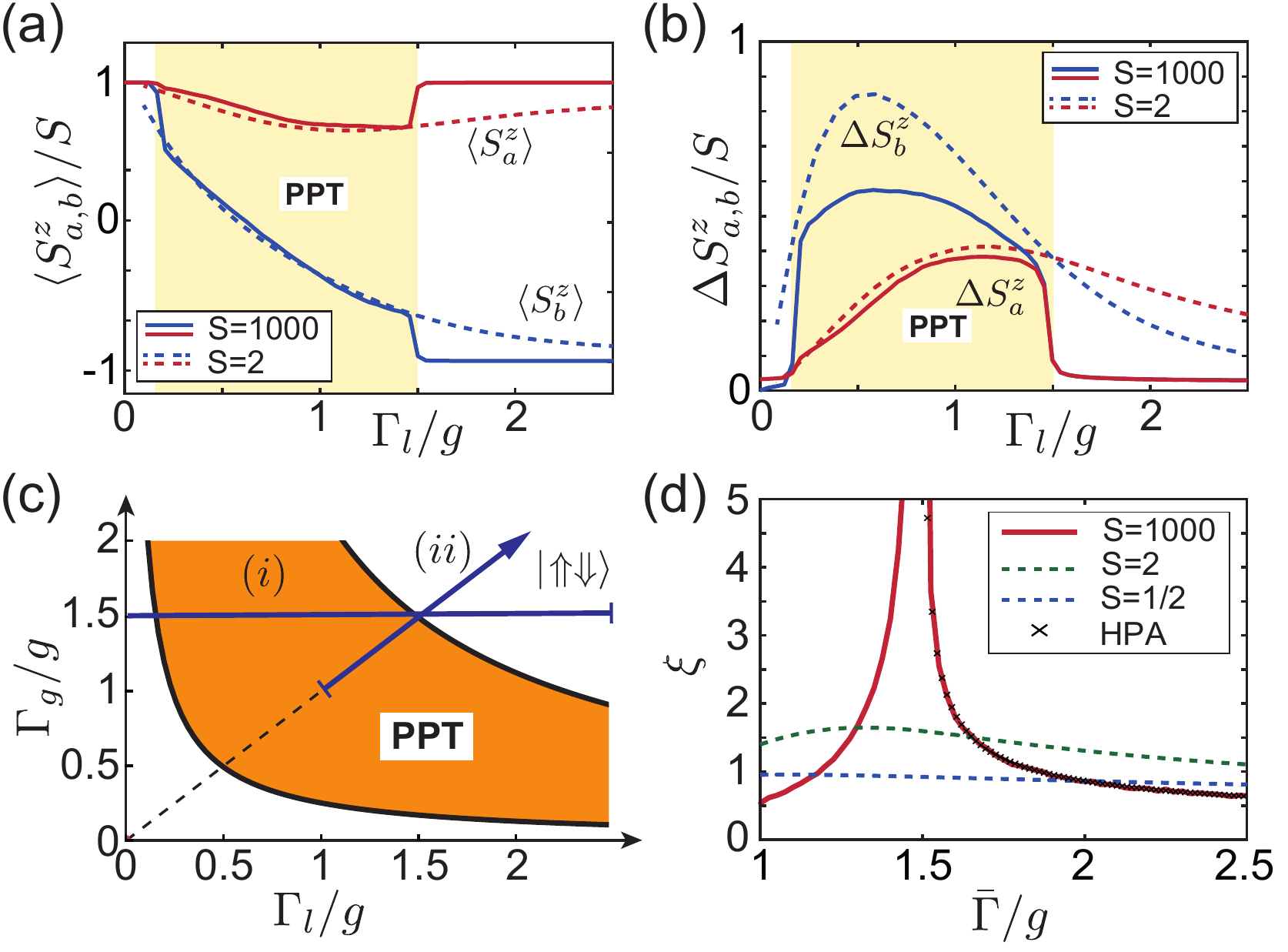}
	\caption{Plot of (a)  the average magnetization $\erw{S^{z}_{a,b}}$ and (b) the magnetization fluctuations, $(\Delta S^{z}_{a,b})^2=\langle (S_{a,b}^z)^2\rangle-\erw{S^{z}_{a,b}}^2$. The two quantities are shown along the path (i) indicated in the sketch of the phase diagram in (c), which shows the extent of the PPT phase for a value of $h/g=0.5$. (d) Plot of the correlation length $\xi$ along the symmetry line $\Gamma_l=\Gamma_g=\bar \Gamma$, i.e. the path (ii) in (c). In (a), (b) and (d) the solid lines represent the results from stochastic simulations based on the TWA for $N=50$ units cells, while the dashed lines have been obtained using iMPO techniques~\cite{Vidal2007,Orus2008}. 
	%{\color{red} (PR: I still don't understand, how the huge fluctuations in (b) fit together with the almost fully polarized spins in (a). Can we do a simple estimate to show that this makes sense? PK: I think the max value is $\Delta S = \sqrt{1-m^2}$ where $m=<S^z>/S$. This maybe contradicts the red line?)}
	}
	\label{Fig3:Chain}
\end{figure}

In Fig.~\ref{Fig3:Chain}(a) and (b) we apply the numerical techniques discussed above to evaluate the dependence of the  average magnetization of each spin and its variance for a fixed $\Gamma_g=1.5 g$ and varying $\Gamma_l$. In the limits $\Gamma_l/g\rightarrow 0$ and $\Gamma_l/g \gg 1$ we recover the FM and AM phases, respectively, which are again characterized by a well-defined magnetization pattern and almost no fluctuations. However, in the extended system, these phases are no longer directly connected. Instead a new intermediate PPT phase appears between the boundaries  $\Gamma_{g} \Gamma_{l}=(g\pm h)^2$.   Although this PPT phase exhibits an imbalanced average magnetization, i.e., $\langle S^z_a\rangle\neq\langle S_b^z\rangle$, it is dominated by large fluctuations similar to the PT phase discussed above. Importantly, this characteristic behavior is no longer restricted to a single line  in parameter space and appears at intermediate values where all dissipation and coherent coupling rates are approximately the same. In the limit $h=g$ the PPT phase completely replaces both FM phases. This shows that the behavior of the lattice systems is considerably different to that of the dimer. For smaller $S$ the boundaries between the phases are much less pronounced, but even in this limit, the three different phases can be clearly distinguished, as can be seen in the results of the iMPO calculations in Fig.~\ref{Fig3:Chain}.

\subsection{Mixed-order transitions}
%In contrast to the case of the spin dimer, the average magnetization changes smoothly when transitioning from the AM into the PPT phase, indicating a second-order phase transition. 
%Note that this continuous variation of $\langle S^z_{a/b}\rangle$ is also found when solving the mean-field equations of motion (see Appendix~\ref{app:MF}) for the same system size. 
In Fig.~\ref{Fig3:Chain}(c) we now take a closer look at the transition between the AM and the PPT phase and evaluate the correlation length $\xi$, as we vary the damping $\bar \Gamma$ across the critical point, $\bar \Gamma_c=g+h$. The correlation length is extracted from an exponential fit of the spin correlation function $\langle S_{a,n}^+S_{a,m}^-\rangle \sim e^{-|n-m|/\xi}$. From our numerical simulations we find that $\xi\sim |\bar \Gamma-\bar \Gamma_c|^{-\nu}$, where  $\nu\simeq 0.5$ in both phases. This behavior would be characteristic for a continuous second-order phase transition associated with the breaking of the $U(1)$ symmetry of our model. However, as shown in Fig.~\ref{Fig3:Chain}(a) the magnetizations $\langle S_{a,b}^z\rangle$ exhibit a rather sharp jump and, as we will argue below, there is no symmetry breaking.

\begin{figure}
	\includegraphics[width=\columnwidth]{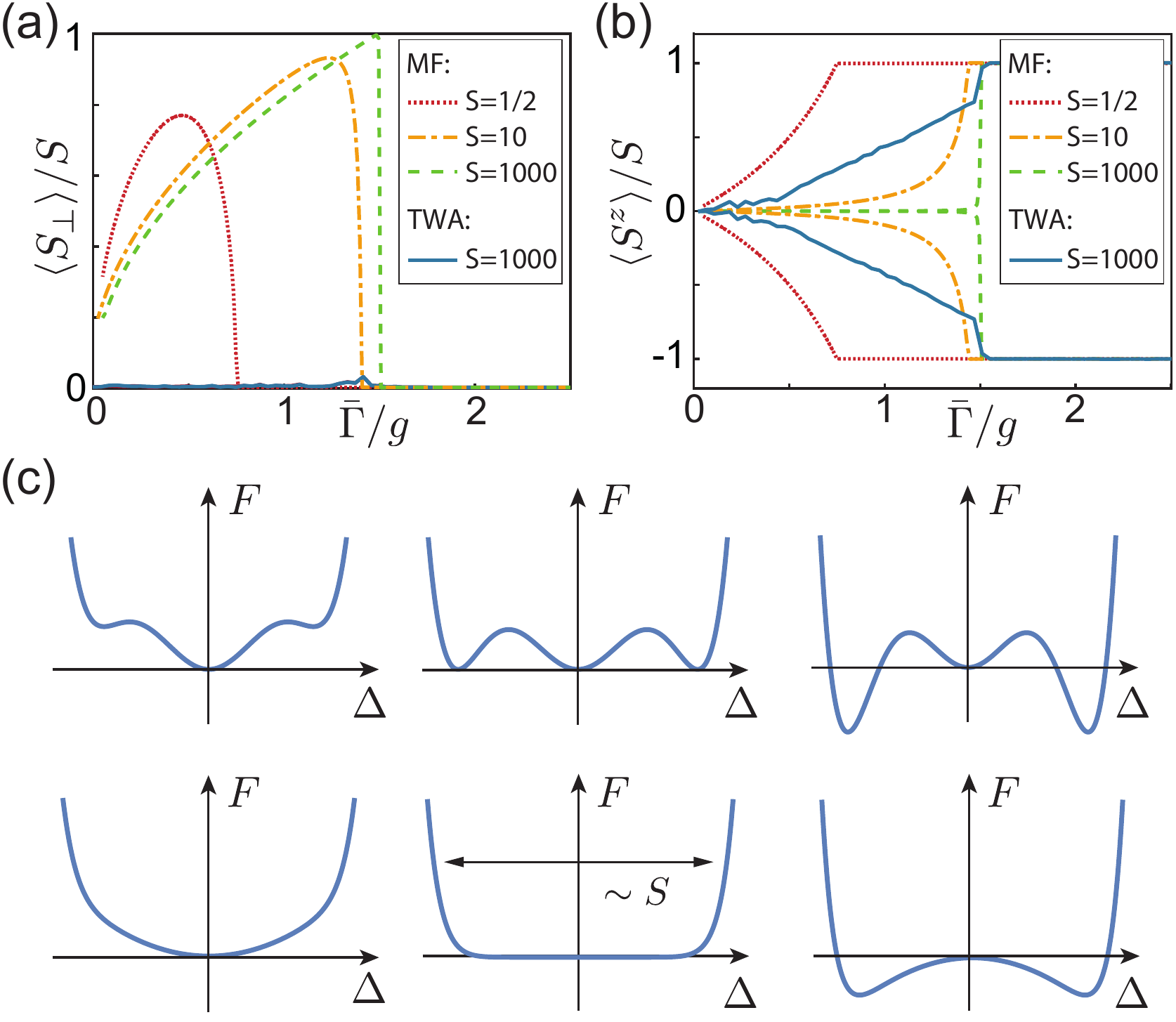}
	\caption{Plot of the steady-state expectation values of (a) the transverse polarization $\langle S_\perp\rangle=\sqrt{\langle S^x_a\rangle^2+\langle S_a^y\rangle^2}$ and (b) the average magnetization $\langle S_z\rangle$ along the symmetry line $\Gamma_{g}=\Gamma_{l}=\bar{\Gamma}$. In both plots the results obtained from a TWA simulation (solid lines) for $S=1000$ and $N=50$ unit cells are compared with the predictions from mean-field (MF) theory (dashed lines) for different spin quantum numbers. (c) Illustration of the difference between a first-order (top row), and a mixed-order (bottom row) phase transition in terms of the usual Landau free energy $F(\Delta)$. The three columns show the variation of the free energy with the order parameter $\Delta$ before (left), at (middle) and after (right) the  transition point.
	}
	\label{Fig5:ChainMF}
\end{figure}

To asses the order of this phase transition we compare in, Fig.~\ref{Fig5:ChainMF}(a) and (b), the results from the full numerical simulation with the predictions from mean-field theory. Mean-field theory shows that while for small spins the transition is indeed continuous, it becomes steeper and steeper with increasing $S$. In the limit $S\rightarrow \infty$ we then obtain a discrete jump in the order parameter $\Delta =  \langle S_a^{-}\rangle$, where for $\delta \Gamma=0$ we obtain the explicit result
\begin{equation}\label{eq:DeltaMF}
\Delta(\bar \Gamma) \simeq  \theta(\bar \Gamma_c-\bar \Gamma)  S\sqrt{\frac{\bar \Gamma}{g+h}}  e^ {i\phi}.
\end{equation}
Here $\theta(x)$ is the Heaviside step function and $\phi$ is an arbitrary phase which breaks the $U(1)$ symmetry~\cite{Sachdev}. In Fig.~\ref{Fig5:ChainMF}(c) and (d) we compare this behavior with two scenarios within the usual Landau free-energy picture of equilibrium phase transitions.  The first case illustrates a first-order transition, where the order parameter jumps from one minimum at $\Delta=0$ to a finite value. If the minima at finite $|\Delta|$ are degenerate, this type of transition can spontaneously break the symmetry, but it will not be associated with a diverging correlation length. The second case depicts a mixed-order transition, where at the transition point the free energy landscape becomes essentially flat. This leads to diverging fluctuations as one approaches the transition point, but also to a jump of the order parameter from $\Delta=0$ to $|\Delta|\sim S$. For small $S$ this picture smoothly connects to the phenomenology of a continuous second-order phase transition.   

Based on this mean-field analogy with conventional Landau theory, we conclude that in the limit of large $S$ the transition from the AM to the PPT phase is most accurately described by a mixed-order phase transition~\cite{Puel2019,Bar2014}. In the exact simulations, the same behavior, namely a jump in the order parameter and a diverging correlation length, is also found for the transition between the FM and the PPT phase, although in this case neither the FM nor the PPT phase are captured by the mean-field equations of motion. For the transition between the two FM phases, the HPA does not predict a diverging correlation length, consistent with a first-order transition as discussed in Sec.~\ref{sec:Dimer}. Of course, this intuitive picture of minimizing an effective potential is very limited and does not take into account the non-equilibrium fluctuations, which, for example, prevent phase-coexistence at the transition point.

\subsection{Absence of symmetry-breaking}

The mean-field result given in Eq.~\eqref{eq:DeltaMF} predicts a breaking of the $U(1)$ symmetry of ME~\eqref{eq:ME}, which is associated with a common rotation of all the spins in the $x$--$y$ plane. However, this symmetry-breaking effect is not observed in our numerical simulations where in all stationary phases $\Delta \simeq0$. As a consequence other expectation values, which are not sensitive to this phase, differ considerably from the mean-field predictions [see Fig.~\ref{Fig5:ChainMF}(a) and (b)]. While expected for small spins,  this observation is very surprising in the limit  $S\to\infty$, where mean-field theory usually becomes exact.

The question of whether or not there is symmetry breaking in the steady state of driven-dissipative systems is actually very subtle, since in the exact steady state all the phases $\phi$ would appear with equal probability and average to zero. Therefore, in the following we use two different numerical approaches to argue that the transition between the AM and PPT phases is inconsistent with our conventional understanding of symmetry-breaking. First, in Fig.~\ref{fig:dynamicsTWMFT}(a) and (b) we show  the results of a CMF simulation (see Appendix~\ref{app:MF}), where the $U(1)$ symmetry is explicitly broken by initializing the spins along a specific direction in the $x$--$y$ plane. Independent of the phase $\phi$, such a state is characterized by a finite value of the transverse spin component
\begin{equation}
\langle S_\perp \rangle =\sqrt{ \langle S^x\rangle^2 +\langle S^y\rangle^2},
\end{equation} 
since it indicates a preferred average direction in the $x$--$y$ plane and hence breaking of the $U(1)$ symmetry. For a cluster size $n_{C}=1$ of one lattice site, which corresponds to the regular mean-field approximation, the broken symmetry is retained in the steady states of the PPT and PT phases. However, as one increases the cluster size, the region with broken symmetry rapidly shrinks and does not considerably grow again when the spin $S$ at each lattice site is increased. This shows that even if the symmetry is explicitly broken by a mean-field ansatz, the system restores the symmetry when the accuracy of the approximation is increased. This behavior must be contrasted to the findings in Refs.~\cite{Rossini2016,Jin2018}. In these references the same scaling analysis correctly predicts the absence of symmetry breaking in 1D, where there is also no phase transition, but supports the existence of a phase with broken symmetry in 2D. Here we find a sharp phase transition but no corresponding symmetry-breaking. 

\begin{figure}
	\includegraphics[width=1\columnwidth]{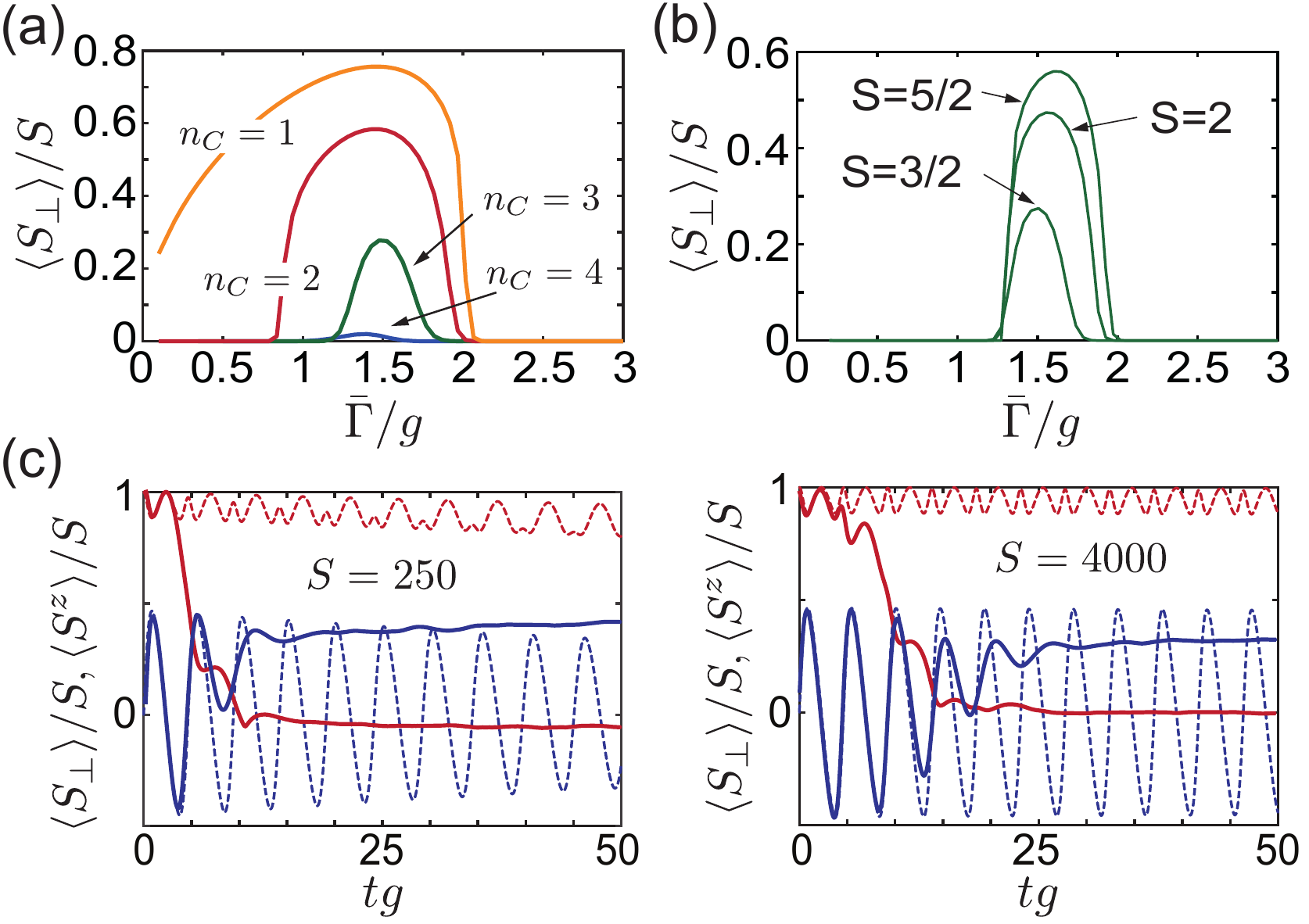}
	\caption{(a) Steady-state expectation value of the transverse polarization $\langle S_\perp\rangle=\sqrt{\langle S^x_a\rangle^2+\langle S_a^y\rangle^2}$ obtained from a CMF simulation with varying cluster size, $n_C$, and $S=3/2$. In (b) the same expectation value is plotted for $n_C=3$ and varying $S$. In both plots $\Gamma_l=\Gamma_g=\bar \Gamma$ and an inter-cell coupling of $h=g$ has been assumed. (c) Dynamics of the spin lattice, which is initially prepared in a symmetry-broken state where all the spins are oriented along the $x$ axis. The solid lines show the TWA results for while the dashed lines are obtained from the mean-field equations of motion. In blue we show $\erw{S_z}$ and in red the perpendicular magnetization $\erw{S_\perp}$. For both plots in (c) the parameters are $\Gamma_{g}=\Gamma_{l}=g$, $h=0.5g$ and $N=50$. }
	\label{fig:dynamicsTWMFT}
\end{figure}

To obtain further evidence for the absence of symmetry breaking in the limit $S\to\infty$, we perform additional dynamical simulations, where the system is initialized in a symmetry-broken state close to the mean-field prediction. We then study the evolution toward the steady state. If the symmetry is broken in the thermodynamic limit we expect that, as we move towards $S\to\infty$, the timescale, $\tau_{\rm sb}$, over which the symmetry is restored should diverge. A prototypical example for such a symmetry-breaking effect is a conventional laser, where the phase diffusion rate decreases inversely with the mean photon number~\cite{GardinerZoller}.

In Fig.~\ref{fig:dynamicsTWMFT}(c) we perform such a numerical experiment on our model in the PPT phase, with $\Gamma_l=\Gamma_g=g $, $h=0.5g$ and $N=50$ unit cells. According to mean-field theory this expectation value stays close to its initial value for the whole duration of the simulation. However, the stochastic simulation, which includes quantum fluctuations from the dissipative processes, shows that this average rapidly approaches zero after a time $\tau_{\rm sb}\sim 10 g^{-1}$, which is also on the order of $\Gamma_{g,l}^{-1}$. Importantly, this time does not considerably increase (by less than a factor of 2), when the spin quantum number is increased by a factor of 16. This gives further evidence to the lack of symmetry breaking in the PPT phase.

We note at this point that the presence of a continuous phase transition without the breaking of the corresponding $Z_2$ symmetry has been previously pointed out for a single-site collective spin model~\cite{Hannukainen2018}, but interpreted as a limiting case of a first-order transition. Since this model also exhibits an infinite-temperature phase, our current analysis suggests an alternative interpretation, namely a purely fluctuation-induced suppression of symmetry breaking.

\section{PT-symmetry breaking in quantum many-body systems}
In the case of the dimer we have already pointed out that ME~\eqref{eq:ME} posses an additional PT symmetry when $\Gamma_l=\Gamma_g$ and that  the PT and AM phases represent the corresponding symmetric and symmetry-broken phases, respectively. Conventionally, PT-symmetry breaking is discussed as a purely \emph{dynamical} effect in systems of coupled classical oscillators with balanced gain and loss~\cite{Ganainy2018}. It is thus an important observation that this mechanism can also influence the stationary states of dissipative quantum systems~\cite{Kepesidis2016,Huber2019,Huber2020} and lead to very unusual transitions between them. Compared to the dimer, an important observation is the appearance of the intermediate PPT phase in the lattice model, which exists over a large parameter range away from the symmetry line.  For these parameters the analogue non-Hermitian oscillator model~\cite{Vazquez-Candanedo2014} has both real and imaginary eigenvalues. Therefore, in this phase the system shares many characteristics of the PT phase, but the symmetry is not fully established.

\begin{figure}
	\includegraphics[width=\columnwidth]{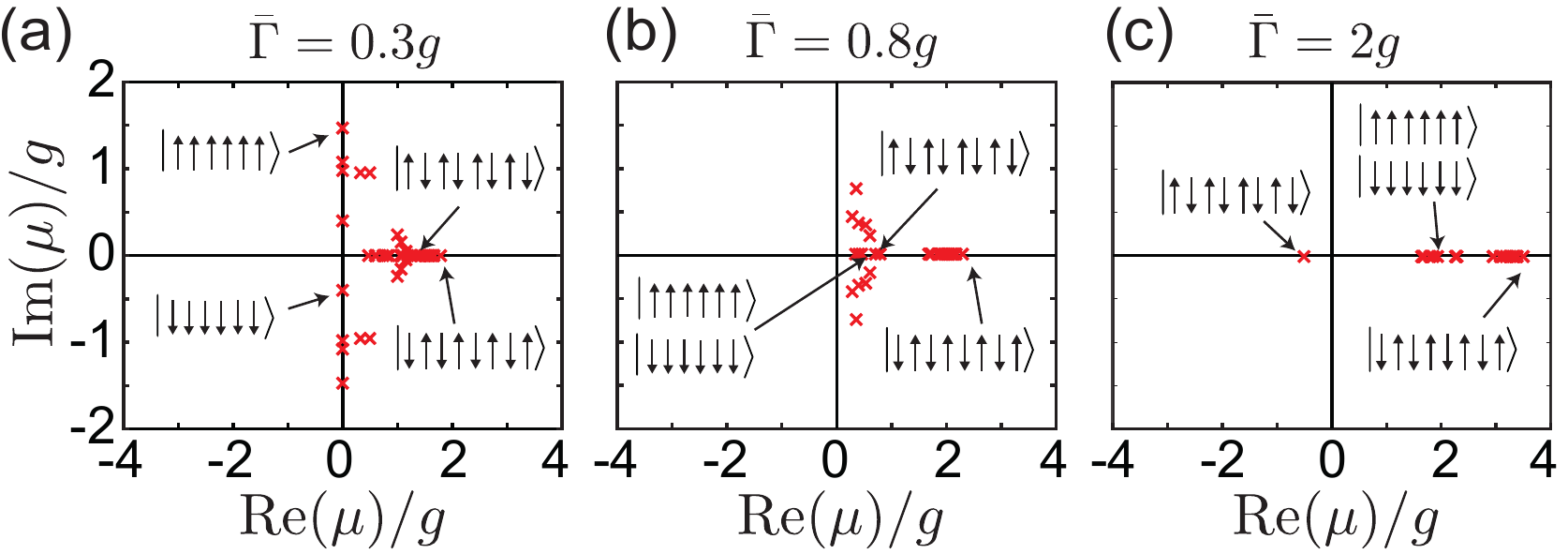}
	\caption{Plot of the eigenvalues $\mu$ of the least stable fluctuation modes for a chain of $N=4$ unit cells, which is initialized in all spin configurations with $\langle S_{a,b}^z\rangle=\pm S$. A few  configurations and the corresponding eigenvalues are shown as examples. }
	\label{Fig4:Eigenvalues}
\end{figure}

To further illustrate this behavior, in Fig.~\ref{Fig4:Eigenvalues} we show the results of a numerical quench experiment. Here, a chain with $N=4$ unit cells is initialized in all $2^{8}$ possible spin configurations with $\langle S_{a,b}^z\rangle=\pm S$.  The successive transient dynamics is characterized by the set of $2^{8}$ complex eigenvalues $\{\mu_{\sigma,i}\}$ of the linearized fluctuation matrix.  For each configuration labeled by $\sigma$, the eigenvalue with the largest real part, representing the least stable fluctuation mode, is shown. For example, in the ordered AM phase, in Fig.~\ref{Fig4:Eigenvalues}(c), there is only a single point with ${\rm Re}(\mu)<0$. This implies that there is only one configuration where all the fluctuations are damped.  All other configurations are rapidly destabilized due to fluctuations that are amplified with rates ${\rm Re}(\mu) \sim \Gamma_{g,l}$. In the PPT phase, Fig.~\ref{Fig4:Eigenvalues}(b), all configurations are unstable, but for a considerable fraction of possible spin orientations the maximal growth rate is very slow, ${\rm Re}(\mu) \ll \Gamma_{g,l}$. Thus, the system transitions slowly  between many metastable orientations, which is reflected in the significant fluctuations observed in this phase.  Another qualitative change is then found in the PT phase,  $\Gamma_l=\Gamma_g<(g-h)$, shown in Fig.~\ref{Fig4:Eigenvalues}(a). Here there are several configurations, where the fluctuations exhibit a purely oscillatory behavior, i.e.,  ${\rm Re}(\mu)=0$, ${\rm Im}(\mu)\sim g$, even in the presence of strong local dissipation. These configurations are neither stable nor unstable, which explains the peculiar properties of this phase. Overall, we see that the pattern of growth rates of spin fluctuations provides a characteristic fingerprint for the different non-equilibrium phases, which can also be used to classify stationary phases of larger lattices, where the exact Liouvillian spectrum is no longer accessible.

\section{Implementation}\label{sec:Implementation}

\begin{figure}
	\includegraphics[width=1\columnwidth]{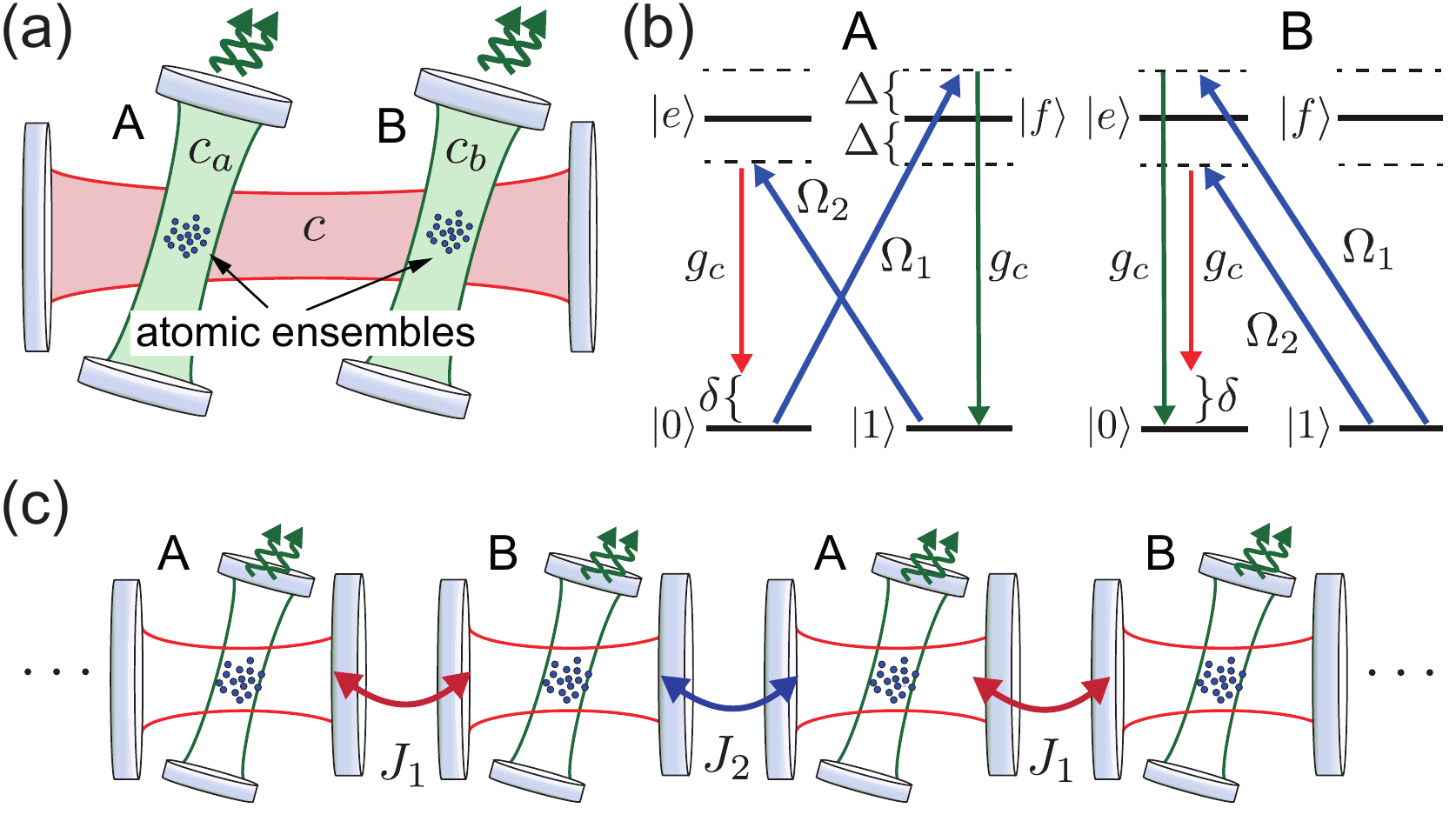}
	\caption{(a) Sketch of a setup for implementing a dissipative spin dimer with gain and loss. The collective cavity mode $c$ is used to mediate coherent interactions between two spin ensembles. The other two cavity modes, $c_a$ and $c_b$, are used to implement collective dissipation channels. (b) Energy level diagram and illustration of the relevant Raman-coupling schemes for realizing effective couplings between the cavity modes and collective spin excitations in the atomic ground states. (c) Generalization to a lattice of tunnel-coupled cavities for implementing the full 1D model considered in this work. See text for more details.}
	\label{fig:implementation}
\end{figure}

While the above analysis is primarily targeted at  a conceptual understanding of non-equilibrium phase transition phenomena, we emphasize that the model in Eq.~\eqref{eq:ME} can be implemented using existing experimental techniques. The basic idea is illustrated in Fig.~\ref{fig:implementation}(a) for a system of cold atoms coupled to multiple optical cavity modes. In this setting, each ensemble contains $N_S$ atoms and is used to encode a collective spin $S=N_S/2$ degree of freedom using two stable atomic ground states $|0\rangle$ and $|1\rangle$, i.e., $S^+= \sum_{i=1}^{N_S}|1\rangle_i\langle 0|$ and $S^z= \sum_{i=1}^{N_S}(|1\rangle_i\langle 1|-|0\rangle_i\langle 0|)/2$. These ground states are coupled via Raman processes involving the excited states $|e\rangle$ and $|f\rangle$ to three different cavity modes with annihilation operators $c$, $c_a$ and $c_b$. The appropriate Raman processes are selected by the choice of detuning and polarization of classical driving fields and are proportional to the atom-cavity coupling strength $g_c$. For simplicity, we assume this coupling to be the same for all modes. For the transitions and detunings indicated in Fig.~\ref{fig:implementation}(b), the resulting effective Hamiltonian for the ground-state spins and the cavity mode is given by~\cite{Dimer2007}
\begin{equation}
\begin{split}
H_{\rm eff}\simeq  \, & \hbar \delta c^\dag c  -\hbar G_c \left[\left(S_a^{-}+S_b^{-}\right)c^\dag   + c \left(S_a^{+}+S_b^{+}\right)\right] \\
&+ \hbar G \left(S_a^{+} c_a^\dag  + c_a S_a^{-}  \right) + G\left(S_b^{-} c_b^\dag  + c_b S_b^{+} \right),
\end{split}
\end{equation}
where we have defined the Raman couplings $G= g_c \Omega_1/ \Delta$ and $G_c=g_c \Omega_2/\Delta$ and $\Omega_{1,2}$ are the Rabi frequencies of the classical driving fields. 

By also including the decay of the cavity modes with rates $\gamma_c$ (for mode $c$) and $\gamma$ (for modes $c_a$ and $c_b$), the dynamics of the full system density operator $\rho_{\rm tot}$ is described by the master equation 
\begin{equation}
\dot \rho_{\rm tot} = -\frac{i}{\hbar} [H_{\rm eff},\rho_{\rm tot}]  + \frac{1}{2}\left( \gamma_c \mathcal{D}[c] +  \gamma \mathcal{D}[c_a] + \gamma_L \mathcal{D}[c_b] \right)\rho_{\rm tot}.
\end{equation}
To proceed we now assume that (i) $|\delta| \gg \gamma_c, G_c$ and (ii) $\gamma \gg G$.  Under these conditions, the coupling to the collective mode $c$ mediates coherent spin-flip interactions, while the resonant coupling to the lossy local modes generates a collective dissipation mechanism. Therefore, after adiabatically eliminating the fast dynamics of the cavity modes we obtain a reduced master equation for the state of the spins, $\rho={\rm Tr}_{c,c_a,c_b}[\rho_{\rm tot}]$. By neglecting common Stark-shift terms for both ensembles, we obtain
\begin{equation}
\dot \rho\simeq -i [g\left(S_a^+S_b^-+S_a^-S_b^+\right),\rho]  + \Gamma_g \mathcal{D}[S_a^+] +  \Gamma_l\mathcal{D}[S_b^-],
\end{equation}
where $g= -G_c^2/ \delta$ and $\Gamma_{g,l}= 2G^2/\gamma$. This is equivalent to ME~\eqref{eq:ME} restricted to a single unit cell. To obtain the full 1D chain, the same schemes can be implemented in an array of coupled cavities, as depicted in Fig.~\ref{fig:implementation}(c), where  the `coherent' mode $c$ from above is replaced by a whole band of the extended modes $c_k$ of the coupled cavity array. As long as the photon-tunneling rates $J_1$ and $J_2$ are small compared to the detuning $\delta$, we obtain approximately nearest-neighbor couplings with  $g\simeq -J_1  G_{c}^2/\delta^2$,  $h= -J_2  G_{c}^2/\delta^2$.

The described setting can be implemented, for example, using cold atoms in multi-mode optical cavities, similar to the experimental setups in Refs.~\cite{Kollar2017,Morales2018,Morales2019}. To realize the full lattice model, one can extend the same techniques to arrays of photonic crystal cavities, as suggested for example in Refs.~\cite{Greentree2006,Hartmann2006}. The coupling of atoms to such nanophotonic structures is currently pursued in several experiments~\cite{Thompson2013,Goban2014}.  In addition, equivalent Raman coupling schemes can be realized with ensembles of solid-state spin qubits, which are coupled magnetically to arrays of microwave resonators~\cite{Zou2014}. This also provides a promising approach for scalable implementations of large-$S$ dissipative spin chains.

\section{Conclusions}
In summary, we have studied the non-equilibrium magnetic phases of a dissipative spin model with gain and loss. These phases and the transitions between them differ in many ways from what is expected for equilibrium systems and from our current understanding of dissipative quantum phase transitions. Specifically, we have found that in this system conventional symmetry breaking is replaced by the dynamical effect of PT-symmetry breaking, which also determines most of the properties of the ordered and disordered phases. 
Note, that by redefining the orientation of all spins on sublattice A, i.e., $S_a^z\rightarrow -S_a^z$, $S_a^+\rightarrow S_a^-$, our model can be mapped onto an XY model with only decay. This model has been studied, for example, in Ref.~\cite{Lee2013} using mean-field theory, where a so-called staggered XY phase with broken $U(1)$ symmetry has been predicted. Our current analysis shows that this phase is more accurately described by a PPT phase without symmetry breaking. This basic example already shows that the effects predicted here are relevant for a much broader class of non-equilibrium models, where such PT-symmetry breaking effects and phase transitions outside the usual framework must be taken into account.

\section*{Acknowledgements}
We dedicate this work to our colleague and dear friend T. Milburn and thank M. Buchold, J.~Keeling, G. Milburn and S. Rotter for stimulating discussions. This work was supported by the Austrian Science Fund (FWF) through Grant No. P32299 (PHONED) and DK CoQuS, Grant No.~W 1210, and through an ESQ fellowship (P.K.) and a  DOC Fellowship (J.H.) from the Austrian Academy of Sciences (\"OAW).

\appendix

\section{Holstein-Primakoff approximation}\label{app:HPApproximation}
In the ordered FM and AM phases and for large $S$ the spins are highly polarized and we can use a HPA~\cite{HolsteinPrimakoff} to linearize  the dynamics of each spin around its mean value on the Bloch sphere.  
%We can use the Holstein-Primakoff approximation to find analytic solutions in the large S limit for different phases. 
Under this approximation the collective spin operators $S^{\pm}$ and $S^z$ are mapped onto a bosonic mode with annihilation operator $c$.  For example, for a spin down state with $\erw{S^z} \approx -S$ we obtain
\begin{align}
S^+&\simeq \sqrt{2S}c^\dagger, & S^-&\simeq \sqrt{2S}c, & S^{z}&=-S+c^\dagger c.
\end{align}
Equivalently, in the opposite limit of a spin up state, where $\erw{S^z} \approx S$, we find
\begin{align}
S^+&\simeq\sqrt{2S}c, & S^-&\simeq \sqrt{2S}c^\dagger, & S^{z}&=S-c^\dagger c.
\end{align}
This approach then allows us to find a description of the system in terms of bosonic modes valid for large $S$ in each of the ordered phases. For example, within the AM phase with all spins pointing up, which we label $\ket{\!\Uparrow\Uparrow}$, we obtain the linearized ME 
\begin{equation}
\dot{\rho}=-\frac{i}{\hbar}[\mathcal{H}_{HPA},\rho]+ \Gamma_{g} \sum_{n=1}^N \mathcal{D}[c_{a,n}]\rho+ \Gamma_{l} \sum_{n=1}^N \mathcal{D}[c_{b,n}^{\dagger}]\rho,
\end{equation}
where $H_{HPA}=\hbar \sum_{n=1}^N (g c_{a,n} c_{b,n}^\dagger +h c_{b,n} c_{a,n+1}^\dagger +{\rm H.c.})$.
%\begin{equation}\label{eq:Hlin}
%\mathcal{H}=\sum_{n=1}^N g c_{a,n} c_{b,n}^\dagger +h c_{b,n} c_{a,n+1}^\dagger +{\rm H.c.}\
%\end{equation}
Here we have introduced the bosonic operators $c_{a,b}$, which describe the left and right spins in each unit cell labeled by $n$. Similar expression are obtained for the other phases, $\ket{\!\Downarrow\Downarrow}$ and $\ket{\!\Uparrow\Downarrow}$.

\subsection{Phase Boundaries}
By assuming periodic boundary conditions, the linearized ME can be solved by changing to Fourier space,
\begin{eqnarray}
c_{a/b,n}=\frac{1}{\sqrt{N}} \sum_{k} e^{i n k} c_{a/b,k} ,
\end{eqnarray}
where the Hamiltonian reads
\begin{equation}\label{eq:HlinFM}
\mathcal{H} = \hbar \sum_{k} g_{k} c_{a,k} c_{b,k}^\dagger + g_{k}^{*} c_{a,k}^\dagger c_{b,k}
\end{equation}
with $g_{k}=g + h e^{i k}$. For the steady-state occupation numbers in $k$-space we then obtain
\begin{eqnarray}
\label{eq:holupup1}
\erw{c_{a,k}^\dagger c_{a,k}}=\frac{\Gamma_{l} \abs{g_k}^2}{(\Gamma_{g}-\Gamma_{l})(\abs{g_k}^2-\Gamma_{g}\Gamma_{l})},\\
\erw{c_{b,k}^\dagger c_{b,k}}=\frac{\Gamma_{l} (\abs{g_k}^2+\Gamma_{g} (\Gamma_{g}-\Gamma_{l}))}{(\Gamma_{g}-\Gamma_{l})(\abs{g_k}^2-\Gamma_{g}\Gamma_{l})},\\
\erw{c_{a,k}^\dagger c_{b,k}}=\frac{i g_{k} \Gamma_{g}\Gamma_{l}}{(\Gamma_{g}-\Gamma_{l})(\abs{g_k}^2-\Gamma_{g}\Gamma_{l})},
\label{eq:holupup2}
\end{eqnarray}
%\subsubsection{Magnetization}
and $\erw{c_{a,k}^\dagger c_{a,k'}}=0$, etc.\ for $k\neq k'$. The corresponding expectation values for each lattice site are given by $\erw{c_{a,n}^\dagger c_{a,n}}=\frac{1}{N} \sum_{k} \erw{c_{a,k}^\dagger c_{a,k}}$ and by approximating this sum by an integral for $N\rightarrow\infty$ we obtain
\begin{eqnarray}
\erw{c_{b,n}^\dagger c_{b,n}}&=& \frac{\Gamma_{l}}{\Gamma_{g}-\Gamma_{l}} \left( 1+ \frac{\Gamma_{g}^2}{C}\right),\\
\erw{c_{a,n}^\dagger c_{b,n}}&=& \frac{i \Gamma_{g} \Gamma_{l}}{2 g (\Gamma_{g}-\Gamma_{l})} \left( 1+ \frac{\Gamma_{g} \Gamma_{l}+g^2-h^2}{C}\right),
\end{eqnarray}
where $C=\sqrt{[(g-h)^2-\Gamma_{g}\Gamma_{l}][(g+h)^2-\Gamma_{g}\Gamma_{l}]}$. Finally, the magnetizations of each of the inequivalent sites are $\erw{S_a^z}=S-	\erw{c_{a,n}^\dagger c_{a,n}}$ and $\erw{S_b^z}=S-\erw{c_{b,n}^\dagger c_{b,n}}$.

These solutions for the occupation numbers only give real numbers when $\Gamma_{g}>\Gamma_{l}$ and $(g-h)^2>\Gamma_{g}\Gamma_{l}$, which shows that the $\ket{\!\Uparrow \Uparrow}$ phase is only stable in these regions of the phase diagram. Note that the same conditions can be obtained from the linear equations of motion for the mean  values $\langle c_{a/b,n}\rangle$.  Equivalent calculations for the $\ket{\!\Downarrow \Downarrow}$ phase give
\begin{eqnarray}
\erw{S_a^z}=-S+\frac{\Gamma_{g}}{\Gamma_{l}-\Gamma_{g}} \left ( 1+ \frac{\Gamma_{l}^2}{C}\right ),\\
\erw{S_b^z}=-S+\frac{\Gamma_{g}}{\Gamma_{l}-\Gamma_{g}} \left ( 1+ \frac{\Gamma_{g}\Gamma_{l}}{C}\right ),
\end{eqnarray}
which are only valid for $\Gamma_{l}>\Gamma_{g}$ and $(g-h)^2>\Gamma_{g}\Gamma_{l}$, where this phase is stable. Finally, for the $\ket{\!\Uparrow \Downarrow}$ phase we find
\begin{eqnarray}
\erw{S_a^z}&=&S-\frac{\Gamma_{l}}{\Gamma_{l}+\Gamma_{g}} \left (-1+ \frac{\Gamma_{g} \Gamma_{l}}{C}\right ),\\
\erw{S_b^z}&=&-S+\frac{\Gamma_{g}}{\Gamma_{l}+\Gamma_{g}} \left (-1+ \frac{\Gamma_{g} \Gamma_{l}}{C}\right ),
\end{eqnarray}
which sets the phase boundary for this phase as $\Gamma_{g} \Gamma_{l} > (g+h)^2$. To obtain the locations of the phase boundaries for the dimer model one may simply set $h=0$ in these expressions.

Note that these results can be generalized in a straightforward manner to higher dimensions and other lattice geometries. For example, in a 2D square lattice we find that all the ordered phases still exist. In this case the antiferromagnetic phase is stable for $\Gamma_{g} \Gamma_l > 4(g +h)^2$, etc.  

%\subsubsection{Boundaries in 2D}
%From the solutions for the $\ket{\Uparrow \Uparrow}$ phase in $k$-space,  Eq.~\eqref{eq:holupup1}-\eqref{eq:holupup2}, it follows that there is stable solution as long as $\abs{g_k}^2 > \Gamma_{g} \Gamma_l$ for all $k$. Equivalently, for the antiferromagnetic phase we find there is a stable solution as long as $\Gamma_{g} \Gamma_l > \abs{g_k}^2$ for all $k$. These results can be straightforwardly extended to higher lattice dimensions. 
%For a 2D square lattice the Hamiltonian reads
%\begin{equation}
%\begin{split}
%\mathcal{H}= \sum_{n,m} &g (a_{n,m} b_{n,m}^\dagger
%+ a_{n,m} b_{n,m+1}^\dagger) \\ +&h (a_{n,m} b_{n+1,m}^\dagger+a_{n+1,m+1} b_{n,m}^\dagger) + h.c.
%\end{split}
%\end{equation}
%In $k$-space we arrive at
%\begin{equation}
%\mathcal{H}= \sum_{k_1,k_2} g_{k_1,k_2} a_{k_1,k_2} b_{k_1,k_2}^\dagger+ h.c.
%\end{equation}
%with $g_{k_1,k_2}= g (1+e^{-i k_{2}})+h (e^{-i k_1}+e^{i (k_1+k_2)}) $.
%Thus, there is an antiferromagnetic phase if $\Gamma_{g} \Gamma_l > \abs{g_{k_1,k_2}}^2$ for all $k_1$ and $k_2$. 

\subsection{Correlation length}

Close to the points where transitions between the different phases occur we see the build-up of long-range correlations in the steady-state density matrix. To quantify these correlations we calculate
\begin{equation}
\begin{split}
\erw{c_{a,n}^\dagger c_{a,n+s}}=\frac{1}{N} \sum_{k} \erw{c_{a,k}^\dagger c_{a,k}} e^{i s k},
%\\
%=\frac{\Gamma_{l}}{\Gamma_{g}-\Gamma_{l}} \left (\frac{\Gamma_{g} \Gamma_{l}}{\sqrt{((g-h)^2-%\Gamma_{g}\Gamma_{l})((g+h)^2-\Gamma_{g}\Gamma_{l})}}\right ) \lambda^{s-1}
\end{split}
\end{equation}
which can be evaluated in the same way as the magnetization above. For example, in the $\ket{\!\Uparrow\Downarrow}$ phase and for $s > 0$ this quantity takes the form 
\begin{equation}
\erw{c_{a,n}^\dagger c_{a,n+s}}=\frac{\Gamma_{l}}{\Gamma_{l}+\Gamma_{g}} \left (\frac{\Gamma_{g} \Gamma_{l}}{C}\right ) \lambda^{s-1},
\end{equation}
where
\begin{equation}
\lambda=\frac{\Gamma_{g}\Gamma_{l}-g^2-h^2-C}{2 g h}.
\end{equation}
This then lets us express the asymptotic form of the spin-spin correlation function as
%for $s>0$ with
%\begin{equation}
%\lambda=\frac{-(g^2+h^2-\Gamma_{g}\Gamma_{l})+\sqrt{((g-h)^2-\Gamma_{g}\Gamma_{l})((g+h)^2-\Gamma_{g}\Gamma_{l})}}{2 g h}
%\end{equation}
%From this follow
\begin{equation}
\abs{\erw{S^{-}_{a,n}S^{+}_{a,n+s}}} \propto e^{-|s|/\xi},
\end{equation}
with the correlation length $\xi=-1/\log(-\lambda)$. 

Close to the phase boundary $\lambda \rightarrow 1$ and the correlation length diverges. 
We can expand around the transition point, $\Gamma_{g} \Gamma_{l}=(g+h)^2$, and find
%\begin{equation}
%\lambda\simeq 1-\sqrt{\frac{\Gamma_{g} \Gamma_{l}-(g+h)^2}{g h}}, %+\O{\Gamma_{g} \Gamma_{l}-(g+h)^2}.
%\end{equation}
%and therefore 
\begin{equation}
\xi=\left (\frac{\Gamma_{g} \Gamma_{l}-(g+h)^2}{g h} \right)^{-1/2}.
\end{equation}
Similar calculations for the other ordered phases show that the critical exponent for the correlation length in this large-spin limit is always $\nu=1/2$.	

\subsection{Purity and Entanglement}

For Gaussian states we can calculate the purity and entanglement negativity from the covariance matrix~\cite{Serafini2014}. Since within the HPA the steady-state is Gaussian we may examine these quantities to understand more about the nature of the phases. This calculation is only analytically tractable in the case of a single dimer, where $h=0$, and so we focus on this case below. For the lattice, the same procedure can be carried out numerically. 

The covariance matrix for the dimer is defined as
\begin{equation}
V_{ij}= \erw{(X_{i}X_{j}+X_{j}X_{i})}/2- \erw{X_{i}} \erw{X_{j}},
\end{equation}
where $X_{1}=(c_{a}+c_{a}^\dagger)$, $X_{2}=i (c_{a}-c_{a}^\dagger)$, $X_{3}=(c_{b}+c_{b}^\dagger)$, $X_{4}=i (c_{b}-c_{b}^\dagger)$. The covariance matrix has the following structure
\begin{equation}
V=
\begin{pmatrix}
V_{A} & V_C\\
V_C^{T} & V_{B}
\end{pmatrix},
\end{equation}
where $V_A$ contains correlations within the first site, $V_B$ those in the second site and $V_C$ the cross-correlations. The covariance matrix of the steady-state can be derived from the linearized master equation in the respective phases. The resulting analytic expression for $V$ is already quite involved and not explicitly shown here.

\subsubsection{Purity}
For a given Gaussian state $\rho$ with co-variance matrix $V$ the purity can be calculated as
\begin{equation}
{\rm Tr}\{\rho^2\}=\frac{1}{\sqrt{\det V}}.
\label{eq:trace}
\end{equation}
In the case of the $\ket{\Uparrow\Downarrow}$ phase the resulting expression for the purity of the steady-state is given by
\begin{equation}
{\rm Tr}\{\rho_0^2\}=\frac{(\Gamma_{g}+\Gamma_{l})^2(\Gamma_{g} \Gamma_{l}-g^2)}{g^2 (\Gamma_{g}-\Gamma_{l})^2+\Gamma_{g} \Gamma_{l} (\Gamma_{g}+\Gamma_{l})^2},
\end{equation}
while in the other two phases $\ket{\Uparrow\Uparrow}$ and  $\ket{\Downarrow\Downarrow}$ we obtain 
\begin{equation}
{\rm Tr}\{\rho_0^2\}=\frac{(\Gamma_{g}-\Gamma_{l})^2 (g^2-\Gamma_{g} \Gamma_{l})}{\Gamma_{g} \Gamma_{l} (\Gamma_{g}-\Gamma_{l})^2+g^2 (\Gamma_{g}+\Gamma_{l})^2}.
\end{equation}
%Similarly, in the $\ket{\Downarrow\Downarrow}$ phase we obtain
%	\begin{equation}
%	\tr(\rho^2)=\frac{(\Gamma_{l}-\Gamma_{g})^2 (g^2-\Gamma_{g} \Gamma_{l})}{\Gamma_{g} \Gamma_{l} (\Gamma_{l}-\Gamma_{g})^2+g^2 (\Gamma_{g}+\Gamma_{l})^2}
%	\end{equation}
%	and finally in the $\ket{\Uparrow\Downarrow}$ phase the purity is given by
%	\begin{equation}
%	\tr(\rho^2)=\frac{(\Gamma_{g}+\Gamma_{l})^2(\Gamma_{g} \Gamma_{l}-g^2)}{g^2 (\Gamma_{g}-\Gamma_{l})+\Gamma_{g} \Gamma_{l} (\Gamma_{g}+\Gamma_{l})^2}.
%	\end{equation}
We see that the purity vanishes at and below the phase boundary and the same behavior is found numerically for larger chains with $h\neq0$.

\subsubsection{Entanglement}
We can calculate the entanglement negativity from the covariance matrix as
\begin{equation}
\mathcal{N}=\frac{1}{2}\left(\frac{1}{\eta}-1\right),
\label{eq:neg}
\end{equation}
where $\eta=\sqrt{\Sigma-\sqrt{\Sigma^2-4 \det V}}/\sqrt{2}$ and $\Sigma=\det V_A + \det V_B - 2 \det V_C$.

By evaluating this expression for both the $\ket{\!\Uparrow\Uparrow}$ and $\ket{\!\Downarrow\Downarrow}$ phases, we obtain a vanishing entanglement, $\mathcal{N}=0$, while the negativity is finite in the $\ket{\!\!\Uparrow\Downarrow}$ phase. This can be understood from the fact that in the former two phases the linearized Hamiltonian contains only excitation-conserving interactions, $\mathcal{H}\sim c_a^\dag c_b+c_a c_b^\dag$, [see Eq.~\eqref{eq:HlinFM}], while in the $\ket{\!\!\Uparrow\Downarrow}$ phase the Hamiltonian creates correlated pairs of excitations, $\mathcal{H}\sim  c_a^\dag c^\dag_b+c_a c_b$. The resulting expression for the  negativity in this phase simplifies along the PT-symmetric line, $\Gamma_{g}=\Gamma_{l}=\bar \Gamma$ to
\begin{equation}
\mathcal{N}=\frac{g}{2 \bar \Gamma}.
\end{equation}
Therefore, the maximal amount of entanglement is reached at the transition point $\bar \Gamma=g$. The same behavior is also found for larger chains when $h\neq0$. Note that within the Holstein-Primakoff approximation a finite amount of entanglement is only found between neighbouring spins.

\section{Mean-field theory}\label{app:MF}
%\subsection{Mean-field theory}\label{Subsec:MF}
From ME~\eqref{eq:ME} we can derive a set of equations of motion for the expectation values of the spin operators,  $\langle S_{a,b}^{x,y,z}\rangle$. Under the mean-field approximation, we factorize all expectation values between two spin operators as $\erw{A B} \rightarrow \erw{A} \erw{B} $ also making the replacement $\langle (S^x)^2+(S^y)^2\rangle =S(S+1)-\langle (S^z)^2\rangle$. We then arrive at the closed but non-linear set of equations, 
\begin{align*}
\erw{\dot S^{x}_{a}}&=-\frac{\Gamma_g}{2S} \erw{S_{a}^{x}}(1+2\erw{S_{a}^{z}})+\frac{(g+h)}{S} \erw{S_{a}^{z}}\erw{S_{b}^{y}},\\
\erw{\dot S^{y}_{a}}&=-\frac{\Gamma_g}{2S} \erw{S_{a}^{y}}(1+2\erw{S_{a}^{z}})- \frac{(g+h)}{S} \erw{S_{a}^{z}}\erw{S_{b}^{x}},\\
\erw{\dot S^{z}_{a}}&= \frac{\Gamma_g}{S} \left[S (S+1)-\erwa{S_{a}^{z}}(\erwa{S_{a}^{z}}+1)\right] \\ 
&+ \frac{(g+h)}{S}( \erw{S_{a}^{y}}\erw{S_{b}^{x}}-\erw{S_{a}^{x}}\erw{S_{b}^{y}}),\\
\erw{\dot S^{x}_{b}}&=-\frac{\Gamma_l}{2S} \erw{S_{b}^{x}}(1-2\erw{S_{b}^{z}})+ \frac{(g+h)}{S} \erw{S_{a}^{y}}\erw{S_{b}^{z}},\\
\erw{\dot S^{y}_{b}}&=-\frac{\Gamma_l}{2S} \erw{S_{b}^{y}}(1-2\erw{S_{b}^{z}})- \frac{(g+h)}{S} \erw{S_{a}^{x}}\erw{S_{b}^{z}},\\
\erw{\dot S^{z}_{b}}&= \frac{\Gamma_l}{S} [S(S+1)-\erw{S_{b}^{z}}(\erw{S_{b}^{z}}-1)]\\
&+\frac{(g+h)}{S} (\erw{S_{a}^{x}}\erw{S_{b}^{y}}-\erw{S_{a}^{y}}\erw{S_{b}^{x}}).
\label{eq:MF}
\end{align*}
Here we have dropped the $n$ subscripts in these equations since under the mean-field approximations each unit cell is identical. These equations can then be readily integrated numerically using standard ODE solvers.

\subsubsection*{Cluster mean-field theory} 
To systematically go beyond the results of the mean-field equations from above, we generalize to the case where all quantum correlations between neighboring sites are included, but a mean-field decoupling is made between these clusters~\cite{Rossini2016}. To achieve this we treat a cluster of $N_C$ unit cells exactly, but factorize the interactions between spins in neighboring clusters. Within this approximation the density operator of the whole chain is replaced by a tensor product of $N/N_C$ smaller density matrices, 
\begin{equation}
\rho\approx  \bigotimes_{\ell=1}^{(N/N_C)} \rho^{(\ell)}_C.  
\end{equation}
Taking the limit $N\to\infty$ allows us to assume translational invariance and hence we set $ \rho^{(\ell)}_C=\rho_C$. The resulting mean-field master equation for $\rho_C$ is given by 
\begin{equation}
\dot{\rho}_C=-\frac{i}{\hbar} [\mathcal{H},\rho_C]+\frac{1}{2 S} \sum_{n=1}^{N_{C}} \left ( \Gamma_{g} \mathcal{D}[S_{a,n}^+]+ \Gamma_{l}  \mathcal{D}[S_{b,n}^-] \right ) \rho_C,
\label{eq:mastereqCMF}
\end{equation}
where
\begin{equation}\label{eq:hamiltonianCMF}
\begin{split}	
\frac{\mathcal{H}}{\hbar} = \frac{1}{S} \sum_{n=1}^{N_{C}-1} & \left[ g \left( S^x_{a,n} S^x_{b,n} + S^y_{a,n} S^y_{b,n} \right)  \right.\\
&+h \left. \left(  S^x_{b,n} S^x_{a,n+1} + S^y_{b,n} S^y_{a,n+1} \right) \right] \\
 &+\frac{h}{S}  \left( \erw{S^x_{b,N_{C}}} S^x_{a,1} +\erw{S^y_{b,N_{C}}} S^y_{a,1} \right) \\
&+ \frac{g}{S} \left( \erw{S^x_{b,1}} S^x_{a,N_{C}} +\erw{S^y_{b,1}} S^y_{a,N_{C}} \right).
\end{split}
\end{equation}
Here, the last two lines of the Hamiltonian account for the mean-field interaction between neighboring clusters. Note that this equation is no longer linear in $\rho_C$ and the evolution of the state and the expectation values must be found self-consistently. 

In our model each unit cell consists of two spin-$S$ systems. This limits the applicability of this method to clusters of size $N_C=1,2$ for even moderate values of $S$. To observe the behavior of the system as the cluster size is increased we thus focus on the symmetric case where $\Gamma_{g}=\Gamma_{l}=\bar{\Gamma}$ and $h=g$. This then allows us to make a unitary transformation which results in a fully translationally model in which the unit cell is a single site. By redefining the spin on every A lattice site as  $S^{z}_{a,n}\rightarrow -S^{z}_{a,n}$, $S^{x}_{a,n}\rightarrow S^{x}_{a,n}$, $S^{y}_{a,n}\rightarrow -S^{y}_{a,n}$, we obtain a model described by the cluster mean-field master equation
\begin{equation}\label{eq:CMF_ME}
\dot{\rho}_C=-\frac{i}{\hbar} [\mathcal{H},\rho_C]+\frac{\bar{\Gamma}}{2 S} \sum_{n=1}^{n_{C}} \mathcal{D}[S_{n}^-] \rho_C,
\end{equation}
with Hamiltonian
\begin{equation}
\begin{split}	
\frac{\mathcal{H}}{\hbar} =& \frac{ g}{S} \sum_{n=1}^{n_{C}-1}  \left ( S^x_{n} S^x_{n+1} - S^y_{n} S^y_{n+1} \right )\\ 
+&\frac{ g}{S}  \left ( \erw{S^x_{n_{C}}} S^x_{1} -\erw{S^y_{n_{C}}} S^y_{1} + \erw{S^x_{b,1}} S^x_{N_{C}} -\erw{S^y_{n_{C}}} S^y_{1} \right) .
\end{split}
\label{eq:hamiltonian}
\end{equation}
This allows us to simulate cluster sizes of $n_C=1,2,3,4$ lattice sites for spin $S\leq 4$ systems.


\begin{thebibliography}{99}
	
%driven-dissipative systems(theory)

\bibitem{Walls1978} D. F. Walls, P. D. Drummond, S. S. Hassan, and H. J. Carmichael, Non-Equilibrium Phase Transitions in Cooperative Atomic Systems, Prog. Theor. Phys. Suppl. {\bf 64}, 307 (1978).

\bibitem{Dimer2007}  F. Dimer, B. Estienne, A. S. Parkins, and H. J. Carmichael, Proposed realization of the Dicke-model quantum phase transition in an optical cavity QED system, Phys. Rev. A {\bf 75}, 013804 (2007).

\bibitem{Morrison2008} S. Morrison and A. S. Parkins, Dynamical Quantum Phase Transitions in the Dissipative Lipkin-Meshkov-Glick Model with Proposed Realization in Optical Cavity QED, Phys. Rev. Lett. {\bf 100}, 040403 (2008).

\bibitem{Prosen2008} T. Prosen and I. Pizor, Quantum Phase Transition in a Far-from-Equilibrium Steady State of an XY Spin Chain, Phys. Rev. Lett. {\bf 101}, 105701 (2008).
% XY chain, with dissipation at the ends

\bibitem{Diehl2008} S. Diehl, A. Micheli, A. Kantian, B. Kraus, H. P. B\"uchler, and P. Zoller, Quantum states and phases in driven open quantum systems with cold atoms, Nat. Phys. {\bf 4}, 878 (2008).


\bibitem{Kessler2012} E. M. Kessler, G. Giedke, A. Imamoglu, S. F. Yelin, M. D. Lukin, and J. I. Cirac, Dissipative phase transition in a central spin system, Phys. Rev. A {\bf 86}, 012116 (2012).

\bibitem{Lee2013} T. E. Lee, S. Gopalakrishnan, and M. D. Lukin, Unconventional Magnetism Via Optical Pumping of Interacting Spin Systems, Phys. Rev. Lett. {\bf 110}, 257204 (2013).

\bibitem{Sieberer2013} L. M. Sieberer, S. D. Huber, E. Altman, and S. Diehl, Dynamical Critical Phenomena in Driven-Dissipative Systems, Phys. Rev. Lett. {\bf 110}, 195301 (2013).

\bibitem{Zou2014} L. Zou, D. Marcos, S. Diehl, S. Putz, J. Schmiedmayer, J. Majer, and P. Rabl, Implementation of the Dicke lattice model in hybrid quantum system arrays, Phys. Rev. Lett. {\bf 113}, 023603 (2014). 

\bibitem{Marzolino2014} U. Marzolino and T. Prosen, Quantum metrology with nonequilibrium steady states of quantum spin chains, Phys. Rev. A {\bf 90}, 062130 (2014).

\bibitem{Carmichael2015} H. J. Carmichael, Breakdown of Photon Blockade: A Dissipative Quantum Phase Transition in Zero Dimensions, Phys. Rev. X {\bf 5}, 031028 (2015).

\bibitem{Weimer2015} H. Weimer, Variational Principle for Steady States of Dissipative Quantum Many-Body Systems, Phys. Rev. Lett. {\bf 114}, 040402 (2015).

\bibitem{Schiro2016} M. Schiro, C. Joshi, M. Bordyuh, R. Fazio, J. Keeling, and H. E. T\"ureci, Exotic attractors of the non-equilibrium Rabi-Hubbard model, Phys. Rev. Lett. {\bf 116}, 143603 (2016). 

\bibitem{Rossini2016} J. Jin, A. Biella, O. Viyuela, L. Mazza, J. Keeling, R. Fazio, and
D. Rossini, Cluster Mean-Field Approach to the Steady-State Phase Diagram of Dissipative Spin Systems, Phys. Rev. X {\bf 6}, 031011 (2016).

\bibitem{Maghrebi2016} M. F. Maghrebi and A. V. Gorshkov, Nonequilibrium Many-Body Steady States via Keldysh Formalism, Phys. Rev. B {\bf 93}, 014307 (2016).

\bibitem{Buchold2017} M. Buchhold, B. Everest, M. Marcuzzi, I. Lesanovsky, and S. Diehl, Nonequilibrium effective field theory for absorbing state phase transitions in driven open quantum spin systems, Phys. Rev. B {\bf 95}, 014308 (2017).

\bibitem{Domokos2017} J. M. Fink, A. Dombi, A. Vukics, A. Wallraff, and P. Domokos, Observation of the Photon-Blockade Breakdown Phase Transition, Phys. Rev. X {\bf 7}, 011012 (2017).


\bibitem{FossFeig2017} M. Foss-Feig, J. T. Young, V. V. Albert, A. V. Gorshkov, and M. F. Maghrebi, Solvable Family of Driven-Dissipative Many-Body Systems, Phys. Rev. Lett. {\bf 119}, 190402 (2017).

\bibitem{Orus2017} A. Kshetrimayum, H. Weimer, and R. Orús, A simple tensor network algorithm for two-dimensional steady states, Nat. Commun. {\bf 8}, 1291 (2017).

\bibitem{Rota2017} R. Rota, F. Storme, N. Bartolo, R. Fazio, and C. Ciuti, Critical behavior of dissipative two-dimensional spin lattices, Phys. Rev. B {\bf 95}, 134431 (2017).

\bibitem{Ciuti2018} F. Vicentini, F. Minganti, R. Rota, G. Orso, and C. Ciuti, Critical slowing down in driven-dissipative Bose-Hubbard lattices, Phys. Rev. A {\bf 97}, 013853 (2018).


\bibitem{Jin2018} J. Jin, A. Biella, O. Viyuela, C. Ciuti, R. Fazio, and D. Rossini, Phase diagram of the dissipative quantum Ising model on a square lattice, Phys. Rev. B {\bf 98}, 241108(R) (2018). 

\bibitem{Minganti2018} F. Minganti, A. Biella, N. Bartolo, and C. Ciuti, Spectral theory of Liouvillians for dissipative phase transitions, Phys. Rev. A {\bf 98}, 042118 (2018).

\bibitem{Hannukainen2018} J. Hannukainen and J. Larson, Dissipation-driven quantum phase transitions and symmetry breaking, Phys. Rev. A {\bf 98}, 042113 (2018).

\bibitem{Roscher2018} D. Roscher, S. Diehl, and M. Buchhold, Phenomenology of first-order dark-state phase transitions, Phys. Rev. A {\bf 98}, 062117 (2018).

\bibitem{Kirton2019} P. Kirton, M. M. Roses, J. Keeling, and E. G. Dalla Torre, Introduction to the Dicke model: from equilibrium to nonequilibrium, and vice versa, Adv. Quantum Technol. {\bf 2}, 1970013 (2019).

\bibitem{Gillman2019} E. Gillman, F. Carollo, and I. Lesanovsky, Numerical Simulation of Critical Dissipative Non-Equilibrium Quantum Systems with an Absorbing State, New J. Phys. {\bf 21}, 093064 (2019). 



\bibitem{Ferreira2019}J. S. Ferreira, and P. Ribeiro, Lipkin-Meshkov-Glick model with Markovian dissipation: A description of a collective spin on a metallic surface, Phys. Rev. B {\bf 100}, 184422 (2019).

\bibitem{Barberena2019} D. Barberena, R. J. Lewis-Swan, J. K. Thompson, and A. M. Rey, Driven-dissipative quantum dynamics in ultra-long-lived dipoles in an optical cavity, Phys. Rev. A {\bf 99}, 053411 (2019).

\bibitem{Puel2019} T. O. Puel, Stefano Chesi, S. Kirchner, and P. Ribeiro, Mixed-Order Symmetry-Breaking Quantum Phase Transition Far from Equilibrium, Phys. Rev. Lett. {\bf 122}, 235701 (2019).

\bibitem{Vicentini2019} F. Vicentini, F. Minganti, A. Biella, G. Orso, and C. Ciuti, Optimal stochastic unraveling of disordered open quantum systems: Application to driven-dissipative photonic lattices, Phys. Rev. A {\bf 99}, 032115 (2019).

\bibitem{Landa2020} H. Landa, M. Schiro, G. Misguich,  Multistability of Driven-Dissipative Quantum Spins, Phys. Rev. Lett. {\bf 124}, 043601 (2020).

\bibitem{Verstraelen2020} W. Verstraelen, and M. Wouters, Classical critical dynamics in quadratically driven Kerr resonators, Phys. Rev. A {\bf 101}, 043826 (2020).



%driven-dissipative experiments
\bibitem{Syassen2008} N. Syassen, D. M. Bauer, M. Lettner, T. Volz, D. Dietze, J. J. Garcia-Ripoll, J. I. Cirac, G. Rempe, and S. D\"urr, Strong Dissipation Inhibits Losses and Induces Correlations in Cold Molecular Gases, Science {\bf 320}, 1329 (2008).

\bibitem{Baumann2010} K. Baumann, C. Guerlin, F. Brennecke, and T. Esslinger, Dicke quantum phase transition with a superfluid gas in an optical cavity, Nature {\bf 464}, 1301 (2010).

\bibitem{Barreiro2011} J. T. Barreiro, M. M\"uller, P. Schindler, D. Nigg, T. Monz, M. Chwalla, M. Hennrich, C. F. Roos, P. Zoller, and R. Blatt, An open-system quantum simulator with trapped ions, Nature {\bf 470}, 486 (2011).

\bibitem{Muller2012} M. M\"uller, S. Diehl, G. Pupillo, and P. Zoller, Engineered Open Systems and Quantum Simulations with Atoms and Ions, Adv. At. Mol. Opt. Phys. {\bf 61}, 1 (2012).
%Review

\bibitem{Safavi-Naini2018}  A. Safavi-Naini, R. J. Lewis-Swan, J. G. Bohnet, M. Garttner, K. A. Gilmore, J. E. Jordan, J. Cohn, J. K. Freericks, A. M. Rey, and J. J. Bollinger, Verification of a Many-Ion Simulator of the Dicke Model Through Slow Quenches across a Phase Transition, Phys. Rev. Lett. {\bf 121}, 040503 (2018).
%non-equilibrium, but dissipation doesn't play an essential role  

\bibitem{Lienhard2018} V. Lienhard, S. de Leseleuc, D. Barredo, T. Lahaye, A. Browaeys, M. Schuler, L.-P. Henry, and A. M. L\"auchli, Observing the Space- and Time-Dependent Growth of Correlations in Dynamically Tuned Synthetic Ising Models with Antiferromagnetic Interactions, Phys. Rev. X {\bf 8}, 021070 (2018).
%non-equilibrium, but dissipation doesn't play an essential role 

\bibitem{Wade2018} C. G. Wade, M. Marcuzzi, E. Levi, J. M. Kondo, I. Lesanovsky, C. S. Adams, and K. J. Weatherill, A terahertz-driven non-equilibrium phase transition in a room temperature atomic vapour, Nature Commun. {\bf 9}, 3567 (2018).

\bibitem{Drummond1980} P. D. Drummond and D. F. Walls, Quantum theory of optical bistability. I. Nonlinear polarisability model, J. Phys. A: Math. Gen.{\bf 13},725 (1980).

\bibitem{Casteels2017} W. Casteels, R. Fazio, and C. Ciuti, Critical dynamical proper-ties of a first-order dissipative phase transition,Phys.Rev.A {\bf 95}, 012128 (2017).

\bibitem{Bartolo2016} N. Bartolo, F. Minganti, W. Casteels, and C. Ciuti, Exact steadystate of a Kerr resonator with one- and two-photon driving and dissipation: Controllable Wigner-function multimodality anddissipative phase transitions, Phys. Rev. A {\bf 94}, 033841 (2016).



%PT Symmetry breaking:
\bibitem{Bender1998} C. M. Bender and S. Boettcher, Real spectra in non-hermitian hamiltonians having PT symmetry, Phys. Rev. Lett. {\bf 80}, 5243 (1998).

\bibitem{Ganainy2018} R. El-Ganainy, K. G. Makris, M. Khajavikhan, Z. H. Musslimani, S. Rotter, and D. N. Christodoulides, Non-Hermitian physics and PT symmetry, Nature Phys. {\bf 14}, 11 (2018).


\bibitem{Huber2020} J. Huber, P. Kirton, S. Rotter, and P. Rabl, Emergence of PT-symmetry breaking in open quantum systems, arXiv:2003.02265 (2020). 




\bibitem{MerminWagner} N. D. Mermin and H. Wagner, Absence of Ferromagnetism or Antiferromagnetism in One- or Two-Dimensional Isotropic Heisenberg Models, Phys. Rev. Lett. {\bf 17}, 1133 (1966).


%imps
\bibitem{Vidal2007}  G. Vidal, Classical Simulation of Infinite-Size Quantum Lattice Systems in One Spatial Dimension, Phys. Rev. Lett. {\bf 98}, 070201 (2007).

\bibitem{Orus2008} R. Orús, and G. Vidal, Infinite time-evolving block decimation algorithm beyond unitary evolution. Phys. Rev. B {\bf 78}, 155117 (2008).	


%TWA
\bibitem{Polkovnikov2010} A. Polkovnikov, Phase space representation of quantum dynamics, Annals of Phys. {\bf 325}, 1790 (2010).
%Review truncated Wigner approximation

\bibitem{Ng2011} R. Ng and  E. S. S\o rensen, Exact real-time dynamics of quantum spin systems using the positive-P representation, J. Phys. A: Math. Theor. {\bf 44}, 065305 (2011).
		
\bibitem{Ng2013} R. Ng, E. S. S\o rensen, and P. Deuar, Simulation of the dynamics of many-body quantum spin systems using phase-space techniques, Phys. Rev. B {\bf 88}, 144304 (2013).
	% spin coherent states, discussion of simulation time
		
\bibitem{Olsen2005} M. K. Olsen, L. I. Plimak, S. Rebic, and A. S. Bradley, Phase-space analysis of bosonic spontaneous emission, Optics Commun. {\bf 254}, 271 (2005).
		%spontaneous decay
		
\bibitem{GardinerZoller} C. W. Gardiner and P. Zoller, {\it Quantum Noise} (Springer, 2000).

\bibitem{HolsteinPrimakoff}  T. Holstein and H. Primakoff, Field dependence of the intrinsic domain magnetization of a ferromagnet, Phys. Rev. {\bf 58}, 1098 (1940).


\bibitem{Huber2021} J. Huber {\it et al.}, in preparation.


\bibitem{Sachdev} S. Sachdev, {\it Quantum Phase Transitions} (Cambridge University
Press, 2001).			

%Mixed order:
\bibitem{Bar2014} A. Bar and D. Mukamel, Mixed-Order Phase Transition in a One-Dimensional Model, Phys. Rev. Lett. {\bf 112}, 015701 (2014).


		

\bibitem{Kepesidis2016} K.  V.  Kepesidis,  T.  J.  Milburn,  J.  Huber,  K.  G.  Makris,  S. Rotter, and P. Rabl, PT-symmetry breaking in the steady state of microscopic gain-loss systems, New J. Phys. {\bf 18}, 095003 (2016).

\bibitem{Huber2019} J. Huber and P. Rabl, Active energy transport and the role of symmetry breaking in microscopic power grids, Phys. Rev. A {\bf 100}, 012129 (2019).



\bibitem{Vazquez-Candanedo2014} O. Vazquez-Candanedo, J. C. Hernandez-Herrejon, F. M. Izrailev, and D. N. Christodoulides, Gain-or loss-induced localization in one-dimensional PT-symmetric tight-binding models, Phys. Rev. A {\bf 89}, 013832 (2014).		
		
		
		

\bibitem{Kollar2017}  A. J. Kollar, A. T. Papageorge, V. D. Vaidya, Y. Guo, J. Keeling, and B. L. Lev, Supermode-density-wave-polariton condensation with a Bose-Einstein condensate in a multimode cavity, Nature Commun. {\bf 8},  14386 (2017).

\bibitem{Morales2018} A. Morales, P. Zupancic, J. Leonard, T. Esslinger, and T. Donner, Coupling two order parameters in a quantum gas, Nature Materials {\bf 17}, 686 (2018).
%two optical cavity modes

\bibitem{Morales2019} A. Morales, D. Dreon, X. Li, A. Baumg\" artner, P. Zupancic, T. Donner, and  T. Esslinger, Two-mode Dicke model from non-degenerate polarization modes,  Phys. Rev. A {\bf 100}, 013816 (2019).

\bibitem{Greentree2006} A. D. Greentree, C. Tahan, J. H. Cole, and L. C. L. Hollenberg, Quantum phase transitions of light, Nature Phys. {\bf 2}, 856 (2006).

\bibitem{Hartmann2006} M. J. Hartmann, F. G. S. L. Brandao, and  M. B. Plenio, Strongly interacting polaritons in coupled arrays of cavities, Nature Phys. {\bf 2}, 849 (2006). 

\bibitem{Thompson2013} J. D. Thompson, T. G. Tiecke, N. P. de Leon, J. Feist, A. V. Akimov, M. Gullans, A. S. Zibrov, V. Vuletic, and M. D. Lukin, Coupling a single trapped atom to a nanoscale optical cavity, Science {\bf 340}, 1202 (2013).

\bibitem{Goban2014} A. Goban, C.-L. Hung, S. P. Yu, J. D. Hood, J. A. Muniz, J. H. Lee, M. J. Martin, A. C. McClung, K. S. Choi, D. E. Chang, O. Painter, and H. J. Kimble, Atom-light interactions in photonic crystals, Nat. Commun. {\bf 5}, 3808 (2014).  



%\bibitem{Zyczkowski} K. Zyczkowski, P. Horodecki, A. Sanpera, and M. Lewenstein, Volume of the set of separable states, Phys. Rev. A {\bf 58}, 883 (1998).

%\bibitem{Vidal} G. Vidal and R. F. Werner, Computable measure of entanglement, Phys. Rev. A {\bf 65}, 032314 (2002).
%

%\bibitem{Meaney2014} C. H. Meaney, H. Nha, T. Duty, and G. J. Milburn, Quantum and classical nonlinear dynamics in a microwave cavity, EPJ QuantumTechnology. {\bf 1}, 1(2014).


		
\bibitem{Serafini2014} A. Serafini, F. Illuminati, M. G. Paris, and S. De Siena, Entanglement and purity of two-mode Gaussian states in noisy channels, Phys. Rev. A, {\bf 69}, 022318 (2004).
		
%\bibitem{Gardiner} C. Gardiner, {\it Stochastic methods} (Springer, Berlin, 2009).










%\bibitem{Rossini2016} J. Jin, A. Biella, O. Viyuela, L. Mazza, J. Keeling, R. Fazio, and
%D. Rossini, Cluster Mean-Field Approach to the Steady-State Phase Diagram of Dissipative Spin Systems, Phys. Rev. X {\bf 6},031011 (2016).


		
%\bibitem{Jin2018} J. Jin, A. Biella, O. Viyuela, C. Ciuti, R. Fazio, and D. Rossini, Phase diagram of the dissipative quantum Ising model on a square lattice, Phys. Rev. B {\bf 98}, 241108 (2018). 






%books



%\bibitem{hofer2017}  P. P. Hofer, M. Perarnau-Llobet, L D. M Miranda, G. Haack,  R. Silva, J. B. Brask, and N. Brunner,  Markovian master equations for quantum thermal machines: local versus global approach,  New J. Phys. {\bf 19}, 123037 (2017).

%\bibitem{Plenio2018} M. T. Mitchison and M. Plenio, Non-additive dissipation in open quantum networks out of equilibrium, New J. Phys. {\bf 20}, 033005 (2018).






%\bibitem{Footnote} Note that there exists a large literature on PT-symmetry breaking effects in non-Hermitian spin models. Since these models do not conserve the norm of the wavefunction, these models do not represent physical meaningful models for dissipative spin systems and cannot be used to predict steady-states. 



% OTHER REFERENCES


%other driven-dissipative systems	

%\bibitem{Diehl2010} S. Diehl, A. Tomadin, A. Micheli, R. Fazio, and P. Zoller, Dynamical Phase Transitions and Instabilities in Open Atomic Many-Body Systems,  Phys. Rev. Lett. {\bf 105}, 015702 (2010).
%
%\bibitem{Prosen2010} T. Prosen and B. Zunkovic, New J. Phys. {\bf 12}, 025016 (2010). 
%%similar to 2008 paper, but also fermions. 
%
%\bibitem{Nagy2011} D. Nagy, G. Szirmai, and P. Domokos, Critical exponent of a quantum-noise-driven phase transition: The open-system Dicke model, Phys. Rev. A {\bf 84}, 043637 (2011).
%
%\bibitem {Oztop2012} B. \"Oztop, M. Bordyuh, \"O. E. M\"ustecaploglu, and H. E. T\"ureci, Excitations of optically driven atomic condensate in a cavity: theory of photodetection measurements, New J. Phys. {\bf 14}, 085011 (2012).
%
%\bibitem{Bohnet2016} J. G. Bohnet, B. C. Sawyer, J. W. Britton, M. L. Wall, A. M. Rey, M. Foss-Feig, and J. J. Bollinger, Science {\bf 352}, 1297 (2016).
%
%\bibitem{Casteels2016} W. Casteels, F. Storme, A. LeBoite, and C. Ciuti, Power Laws in the Dynamic Hysteresis of Quantum Nonlinear Photonic Resonators, Phys. Rev. A {\bf 93}, 033824 (2016).
%
%\bibitem{Hwang2018} M.-J. Hwang, P. Rabl, and M. B. Plenio, Dissipative Phase Transition in the Open Quantum Rabi Model, Phys. Rev. A {\bf 97}, 013825 (2018). 
%
%\bibitem{CarusottoRMP2013} I. Carusotto and C. Ciuti, Rev. Mod. Phys. {\bf 85}, 299 (2013).
%
%%large S
%\bibitem{Morrison2008} S. Morrison and A. S. Parkins, Dissipation-driven quantum phase transitions in collective spin systems, J. Phys. B: At. Mol. Opt. Phys. {\bf 41}, 195502 (2008).
%
%\bibitem{Larson2018} J. Hannukainen and J. Larson, Dissipation-driven quantum phase transitions and symmetry breaking, Phys. Rev. A {\bf 98}, 042113 (2018).


%\bibitem{Horstmann2013} B. Horstmann, J. I. Cirac, and G. Giedke, Noise-driven dynamics and phase transitions in fermionic systems, Phys. Rev. A {\bf 87}, 012108 (2013).

%\bibitem{Verstraete2009} F. Verstraete, M. M. Wolf, and J. I. Cirac, Quantum computation and quantum-state engineering driven by dissipation, Nat. Phys. {\bf 5}, 633 (2009).

%XYZ
%\bibitem{Lee2011}  T. E. Lee, H. Häffner, and M. C. Cross, Antiferromagnetic Phase Transition in a Nonequilibrium Lattice of Rydberg Atoms, Phys. Rev. A {\bf 84}, 031402 (2011).

%experiments
%\bibitem{Monroe2010} K. Kim, M. S. Chang, S. Korenblit, R. Islam, E. E. Edwards, J. K. Freericks, G. D. Lin, L. M. Duan, and C. Monroe, Quantum simulation of frustrated Ising spins with trapped ions, Nature {\bf 465}, 590 (2010).
%\bibitem{Blatt2011} J. T. Barreiro, M. M\"uller, P. Schindler, D. Nigg, T. Monz, M. Chwalla, M. Hennrich, C. F. Roos, P. Zoller, and R. Blatt, An open-system quantum simulator with trapped ions, Nature {\bf 470}, 486 (2011).
%\bibitem{Bloch2015} M. Schreiber, S. S. Hodgman, P. Bordia, H. P. L\"uschen, M. H. Fischer, R. Vosk, E. Altman, U. Schneider, and I. Bloch, Observation of many-body localization of interacting fermions in a quasi-random optical lattice, Science {\bf 349}, 842 (2015)
%\bibitem{Houck2012} A. A. Houck, H. E. Tureci, and J. Koch, On-chip quantum simulation with superconducting circuits, Nat. Phys. {\bf 8}, 292 (2012).
%\bibitem{Houck2017} M. Fitzpatrick, N. M. Sundaresan, A. C. Y. Li, J. Koch, and
%A. A. Houck, Observation of a Dissipative Phase Transition in a One-Dimensional Circuit QED Lattice, Phys. Rev. X {\bf 7}, 011016 (2017).

%\bibitem{Lukin2017} H. Bernien, S. Schwartz, A. Keesling, H. Levine, A. Omran, H. Pichler, S. Choi, A. S. Zibrov, M. Endres, M. Greiner, V. Vuletic, and M. D. Lukin, Probing many-body dynamics on a 51-atom quantum simulator, Nature (London) {\bf 551}, 579 (2017).
%\bibitem{Lienhard} V. Lienhard, S. de Leseleuc, D. Barredo, T. Lahaye, A. Browaeys, M. Schuler, L.-P. Henry, and A. M. L\"auchli, Observing the Space- and Time-Dependent Growth of Correlations in Dynamically Tuned Synthetic Ising Models with Antiferromagnetic Interactions, Phys. Rev. X {\bf 8}, 021070 (2018).
%\bibitem{Fink2018}  T. Fink, A. Schade, S. Höfling, C. Schneider, and A. Imamoglu, Signatures of a dissipative phase transition in photon correlation measurements, Nat. Phys. {\bf 14}, 365 (2018).

%\bibitem{Bloch2008} I. Bloch, J. Dalibard, and W. Zwerger, Many-body physics with ultracold gases, Rev. Mod. Phys. {\bf 80}, 885 (2008).


%non-Hermitian quantum many-body models:
%\bibitem{Lee2014} T. E. Lee and C.-K. Chan, Heralded Magnetism in Non-Hermitian Atomic Systems, Phys. Rev. X {\bf 4}, 041001 (2014).
%
%%Ryusuke Hamazaki, Kohei Kawabata, Masahito Ueda, Non-Hermitian Many-Body Localization,arXiv:1811.11319 
%
%\bibitem{Ashida2017} Y. Ashida, S. Furukawa, and M. Ueda, Parity-time-symmetric quantum critical phenomena,  Nature Commun. {\bf 8}, 15791 (2017).
%%non-Hermitian sineGordon and effective spin models. 




\end{thebibliography}
\end{document}